\documentclass[fleqn,usenatbib]{mnras}

\usepackage{newtxtext,newtxmath}

\usepackage[T1]{fontenc}

\DeclareRobustCommand{\VAN}[3]{#2}
\let\VANthebibliography\thebibliography
\def\thebibliography{\DeclareRobustCommand{\VAN}[3]{##3}\VANthebibliography}

\usepackage{graphicx}	
\usepackage{amsmath}	
\usepackage[dvipsnames]{xcolor}

\usepackage{subcaption}
\newcommand{\R}{\mathbb R}
\usepackage{changes} 
\usepackage[dvipsnames]{xcolor} 
\usepackage{float}

\title[Investigating Cosmological GAN Emulators]{Investigating Cosmological GAN Emulators Using Latent Space Interpolation}

\author[A. Tamosiunas et al.]{
Andrius Tamosiunas,$^{1,3}$\thanks{E-mail: andrius.tamosiunas@nottingham.ac.uk }
Hans A. Winther,$^{2}$
Kazuya Koyama,$^{3}$
David J. Bacon,$^{3}$
Robert C. Nichol$^{3}$ 
\newauthor and Ben Mawdsley$^{3}$
\\
$^{1}$School of Physics and Astronomy, University of Nottingham, University Park ,Nottingham, NG7 2RD, United Kingdom\\
$^{2}$Institute of Theoretical Astrophysics, University of Oslo, Svein Rosselands hus, Blindern campus Sem Saelandsvei 13 0371 Oslo, Norway\\
$^{3}$Institute of Cosmology and Gravitation, University of Portsmouth,Dennis Sciama Building, Burnaby Road, Portsmouth, PO1 3FX, United Kingdom
}

\date{Accepted XXX. Received YYY; in original form ZZZ}

\pubyear{2020}

\begin{document}
\label{firstpage}
\pagerange{\pageref{firstpage}--\pageref{lastpage}}
\maketitle

\begin{abstract}
Generative adversarial networks (GANs) have been recently applied as a novel emulation technique for large scale structure simulations. Recent results show that GANs can be used as a fast and efficient emulator for producing novel weak lensing convergence maps as well as cosmic web data in 2-D and 3-D. However, like any algorithm, the GAN approach comes with a set of limitations, such as an unstable training procedure, inherent randomness of the produced outputs and difficulties when training the algorithm on multiple datasets. In this work we employ a number of techniques commonly used in the machine learning literature to address the mentioned limitations. Specifically, we train a GAN to produce weak lensing convergence maps and dark matter overdensity field data for multiple redshifts, cosmological parameters and modified gravity models. In addition, we train a GAN using the newest Illustris data to emulate dark matter, gas and internal energy distribution data simultaneously. Finally, we apply the technique of latent space interpolation as a tool for understanding the feature space of the GAN algorithm. We show that the latent space interpolation procedure allows the generation of outputs with intermediate cosmological parameters that were not included in the training data. Our results indicate a 1-20\% difference between the power spectra of the GAN-produced and the test data samples depending on the dataset used and whether Gaussian smoothing was applied. Similarly, the Minkowski functional analysis indicates a good agreement between the emulated and the real images for most of the studied datasets.

\end{abstract}




\section{Introduction}

\label{sec:intro}
\raggedbottom

In the era of precision cosmology an important tool for studying the evolution of large scale structure is \textit{N}-body simulations. Such simulations evolve a large number of particles under the influence of gravity (and possibly other forces) throughout cosmic time and allow detailed studies of the non-linear structure formation. Modern cosmological simulations are highly realistic and extremely complex and may include galaxy evolution, feedback processes, massive neutrinos, weak lensing and many other effects. Such complexity however comes at a price in terms of computational resources and large simulations may take several days or even weeks to run. In addition, to fully account for galaxy formation and other effects various simplification schemes and semi-analytical models are required. To address these issues a variety of emulation techniques have been discussed in the literature \citep{Kwan2015, winther2019, knabenhans2019}. In light of upcoming surveys like Euclid, such emulators will be an invaluable tool for producing mock data quickly and efficiently.

Lately, machine learning techniques have been explored as a valuable tool in cosmology, with applications ranging widely from cosmological parameter extraction from observational data to Supernovae classification \citep{ntampaka2019,merten2019}. Machine learning techniques have also been applied as an alternative to the traditional emulation methods. For instance, deep learning has been used to accurately predict non-linear structure formation \citep{He2019}. Similarly GANs and variational autoencoders (VAEs) have been used to produce novel realistic cosmic web 2-D projections, weak lensing maps and to perform dark energy model selection \citep{Kingma2013,goodfellow2014,rodriguez2018, Mustafa2019, Li2019, Ullmo2020}. In addition the GAN approach has also been used to produce realistic cosmic microwave background temperature anisotropy 2-D patches as well as deep field astronomical images \citep{Mishra2019, Smith2019}. Finally, generating full 3-D cosmic web data has been discussed in \cite{perraudin2019, ramanah2020}. The cited works show that GANs are capable of reproducing a variety of cosmological simulation outputs efficiently and with high accuracy. 

However, certain challenges remain: the training process of the GAN algorithm is complicated and prone to failure and producing full scale 3-D results is computationally expensive. A common problem when training GANs is \textit{mode collapse}, when the generator neural network overpowers the discriminator and gets stuck in producing a small sample of identical outputs. Mode collapse can be addressed in multiple ways -- modern GAN architectures introduce label flipping or use different loss functions, such as Wasserstein distance, which has been shown to reduce the probability of mode collapse \citep{arjovsky2017}. 

In addition, GANs and other neural network based approaches suffer from being \textit{black box} algorithms -- i.e. it is usually difficult to determine the specific features that are deemed important by the algorithm during the training procedure. This issue is of special importance when training GANs on scientific data as such information could hold important clues about the non-trivial correlations in the dataset. For instance, a GAN trained on weak lensing convergence maps with different cosmological parameters, could contain important information about the effects of cosmology on weak lensing that is not captured by the usual summary statistics.

In this paper we address some of these issues and present our results on extending some of the currently existing GAN algorithms. In particular, we use a modified version of the \textit{cosmoGAN} algorithm (introduced in \cite{Mustafa2019}) to produce weak lensing convergence maps and 2-D cosmic web projections of different redshifts and multiple cosmologies. In addition, the GAN is trained on dark matter, gas and internal energy data simultaneously. Furthermore, we explore techniques from contemporary research in the field of deep learning, such as latent space interpolation, as a way to control the outputs of the algorithm and as a tool to explore the structure of the latent space. Finally, we apply the latent space interpolation techniques for producing data of unseen cosmological parameters.

Similar issues have been recently explored in \cite{rodriguez2018, Mustafa2019, Perraudin2020}. In \cite{Perraudin2020} in particular, the authors employ a conditional GAN architecture to produce weak lensing convergence maps with different cosmological parameters \cite{Mirza2014}. The key feature of the conditional GAN architecture is that in addition to the latent space vector input one adds a parameter vector, which allows one to specify the cosmological parameters directly. This feature allows direct control of the algorithm outputs. The main difference of our approach is that the DCGAN \citep{Radford2015} architecture described in this work does not allow direct control of the outputs. This introduces extra difficulties when training the GAN algorithm on dataset that consists of data with multiple cosmological parameters. However, we explore multiple alternative approaches based on latent space interpolation. In doing so we also explore the structure of the latent space produced during the training procedure, which is the first step towards tackling the black box problem of the GAN algorithm.

Finally, we discuss GANs in the framework of Riemannian geometry in order to put our problem on a more theoretical footing and to explore the feature space learnt by the algorithm.

\section{Generative Adversarial Networks}
\subsection{The Algorithm}

GANs, first introduced in a now seminal paper \cite{goodfellow2014}, are a system of neural networks that are trained adversarially. In particular, a GAN consists of a \textit{generator} -- a neural network responsible for producing data from random noise and a \textit{discriminator}, which is responsible for evaluating the produced data against the training set. The two neural networks compete in an adversarial fashion during the training process -- the generator is optimized to produce realistic datasets statistically identical to the training data and hence to fool the discriminator. Mathematically, such an optimization corresponds to minimizing the following cost function: 

\begin{equation}
\begin{aligned}
 \min_{G_{\phi}} \max_{D_{\theta}}J(D_{\theta},G_{\phi}) =& - \mathbb{E}_{X \sim \mathbb{P}_{r}}\log(D_{\theta}(X)) \\
 & - \mathbb{E}_{Z \sim \mathbb{P}_{g}}\log(1 - D_{\theta}(G_{\phi}(Z))),  
 \end{aligned}
\end{equation}

\noindent where $\mathbb{E}$ refers to the expectation function, $D_{\theta}$ to the discriminator with weights $\theta$, $G_{\phi}$ to the generator with weights $\phi$, $\mathbb{P}_{r}$ to the distribution of the data we are aiming for, $\mathbb{P}_{g}$ to the generated distribution, $X$ to the data (real or generated) analyzed by the discriminator and $Z$ to the random noise vector input to the generator.

Such an optimization procedure is a nice example of game theory where the two agents (the generator and the discriminator) compete in a two player zero sum game and adjust their \textit{strategies} (neural network weights) based on the common cost function. In case of perfect convergence, the GAN would reach Nash equilibrium, i.e. the generator and the discriminator would reach optimal configurations (optimal sets of weights). In practice, however, reaching convergence is difficult and the training procedure is often unstable and prone to mode collapse \citep{farnia2020}. 

The two neural networks, the discriminator and the generator, have two different training procedures. In particular, the discriminator classifies the datasets into real (coming from the training dataset) or fake (produced by the generator) and is penalized for misclassification via the discriminator loss term. The discriminator weights are updated through backpropagation as usual \citep{Linnainmaa1976, Rumelhart1986}. The generator, on the other hand, samples random noise, produces an image, gets the classification of that image from the discriminator and updates its weights accordingly via backpropagation using the generator loss function term. The full training procedure is done by alternating between the discriminator and the generator training cycles.    

\begin{figure}
  \centering
    \includegraphics[width=1.05\linewidth]{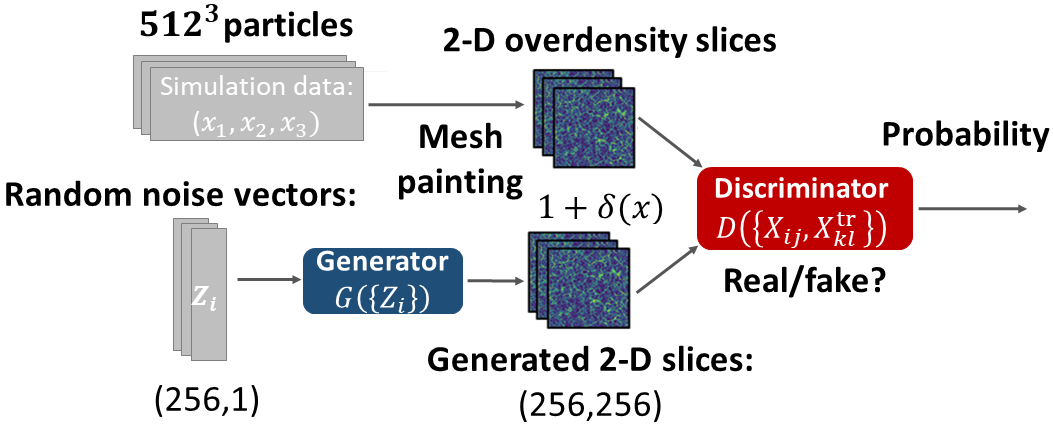}
    \caption{The pipeline of training a GAN on 2-D DM-only cosmic web slices. A number of $512^{3}$ particle simulation boxes are mesh-painted and sliced to produce the $256 \times 256$ px cosmic web slice training dataset. The training dataset is then used to train the system of the generator and the discriminator neural networks. Once the training procedure is finished, the generator can be used to generate novel 2-D cosmic web slices out of random Gaussian noise vectors. A nearly analogous procedure is used to train the weak lensing convergence maps, with the main difference being that an extra step of ray-tracing is required in order to produce the training dataset.}
    \label{figure 1}
\end{figure}

Assuming the adversarial training is successful, the generator $G_{\phi}(Z)$ can then be used separately for producing realistic synthetic data from a randomized input vectors $Z$. Fig. \ref{figure 1} lays out the pipeline for using a GAN to generate DM-only cosmic web slice data\footnote{A note on the used terminology: cosmic web slices in this work refer to the 2-D overdensity field projections generated by slicing full 3-D overdensity field data from \textit{N}-body simulations. Such slices are then used to train the GAN.}.

\subsection{Latent Space Interpolation}

The generator neural network with its multi-layered structure can be represented mathematically as a function composition: 

\begin{equation}
    G_{\phi}(Z_{i}) = g^{1} \circ g^{2} \circ ... \circ g^{n}
    \quad\text{with}\quad
    g_{k}^{i}(y^{i}) = S(W_{k}^{i}y^{i} + b^{i}),
\end{equation}

\noindent where each layer $g^{i}$ maps from an input $y^{i}$ to an output as shown above. Here $G_{\phi}(Z_{i})$ is the generator neural network, $Z_{i}$ is the random input vector, $S(y)$ is a non-linear activation function, $W_{k}^{i}$ is the weight matrix and $b^{i}$ is the bias term. The aim of the training procedure is to find an optimal weight matrix $W$ (along with the bias terms), which maps the input to the wanted output.    

If the training procedure is successful, the generator $G_{\phi}(Z_{i})$ learns to map the values of a random vector $Z_{i}$ to the values of a statistically realistic 2-D array representing the output $X_{jk}$ (a cosmic web slice or a convergence map in our case). This can be viewed as mapping from a low-dimensional \textit{latent} space $Z \subseteq \R^{d}$ to a higher-dimensional data (pixel) space $X \subseteq \R^{D}$ (for more details see \cite{shao2017} and appendix \ref{appendix:riemannian_geometry}). For a generator neural network $d \ll D$ (in our case $d = 256$ or $64$, while $D = 256^{2}$).

\begin{figure*}
  \centering
    \includegraphics[width=0.6\textwidth]{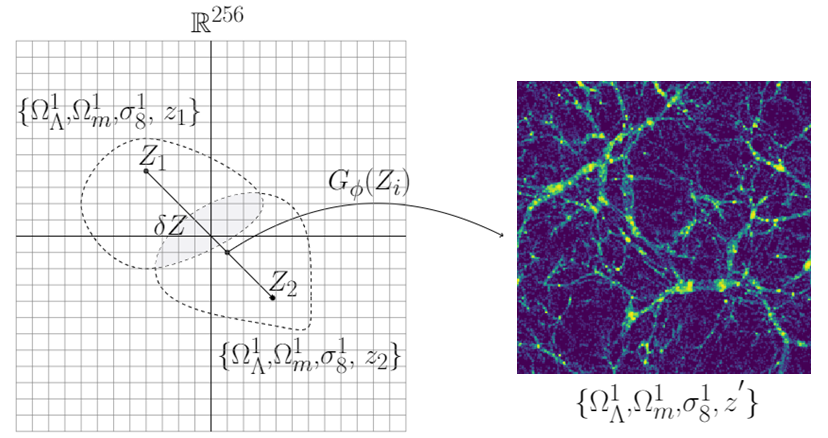}
    \caption{Illustration of the latent space interpolation procedure. Training the GAN algorithm on the cosmic web slices of two different redshifts encodes two different clusters in the latent space (which is a subset of a 256-dimensional space, i.e. the size of the random noise input vector). Sampling a point from the line connecting two input points $Z_{1}$ and $Z_{2}$ in this space produces an output with redshift  $z^{'}$. As we will see, in the case of our dataset with $z_{1} = 1.0$ and $z_{2} = 0.0$, several points near the centre of this line correspond to outputs approximately emulating $z^{'} \approx 0.5$. }
    \label{figure 2}
\end{figure*}

The generator network has a number of interesting properties. In particular, during the training procedure it maps clusters in the $Z$ space to the clusters in the $X$ space. Hence, if we treat the random input vectors $Z_{i}$ as points in a $d$-dimensional space, we can interpolate between multiple input vectors and produce a transition between the corresponding outputs. In particular, if we choose two input vectors $Z_{1}$ and $Z_{2}$ and find a line connecting them, sampling intermediate input points $Z_{i}$ along that line leads to a set of outputs that correspond to an almost smooth transition between outputs $X_{1}$ and $X_{2}$. As an example, if we train the generator to produce cosmic web slices of two different redshifts, we can produce a set of outputs corresponding to a transition between those two redshifts by linearly interpolating between the input vectors $Z_{1}$ and $Z_{2}$ (see fig. \ref{figure 2}). More concretely, if we train the algorithm on cosmic web slices of redshifts $\{0.0,1.0\}$, somewhere between the two input vectors, one can find a point $Z^{'}$, which produces an output that has a matter power spectrum approximately corresponding to a redshift  $z^{'}\approx 0.5$ (see fig. \ref{figure10} and appendix \ref{appendix:latent_space_clustering}). This is fascinating given that the training dataset did not include intermediate redshift data. Here it is important to note that such an interpolation procedure does not necessarily produce a perfectly smooth transition in the data space, i.e. the produced outputs corresponding to the latent space vectors $Z_{i}$ between $Z_{1}$ and $Z_{2}$ are not always realistic (in terms of the matter power spectrum and other statistics; see fig. \ref{figure11} and section \ref{section:latent_results} for further details). Also, one might naively think that the point $Z^{'}$ lies in the middle of the line connecting $Z_{1}$ and $Z_{2}$, but in general we found it not to be the case (as the middle of the mentioned line does not necessary correspond to the middle between $X_{1}$ and $X_{2}$ in the data space, which is known to be non-Euclidean (see appendix \ref{appendix:riemannian_geometry})). In this work we investigate whether the latent space interpolation procedure can be used to map between outputs of different redshifts and cosmologies and whether the produced datasets are physically realistic. 

The latent space interpolation technique was performed by randomly choosing two input vectors $Z_{1}$ and $Z_{2}$, finding the line connecting the two points in the 256 (64)-dimensional space (256 (64) is the size of $Z_{1}$ and $Z_{2}$) and then sampling 64 equally spaced points along that line. The outputs of the generator neural network of those intermediate input points $G_{\phi}(Z_{int})$ then correspond to cosmic web slices and weak lensing maps that represent a transition between the two outputs $G_{\phi}(Z_{1})$ and $G_{\phi}(Z_{2})$.

In order to perform linear latent space interpolation it is crucial to have the ability to distinguish between different data classes produced by the GAN (e.g. cosmic web slices of different redshifts). To resolve this problem we employed a combination of the usual summary statistics like the power spectrum and the Minkowski functionals along with a machine learning algorithm. In particular, we tested using gradient boosted decision trees for distinguishing the different classes of datasets produced by the GAN \citep{chen2016_xgboost} (see section \ref{section:latent_results} for a further discussion of the chosen approach).

\section{Datasets and the Training Procedure}

\subsection{Weak Lensing Convergence Map Data}

Gravitational potentials influence the path of photons in such a way that they introduce coherent distortions in the apparent shape (shear) and position of light sources. Weak gravitional lensing introduces ellipticity changes in objects of the order of $\approx$ 1\% and can be measured across the sky, meaning that maps of the lensing distortion of objects can be made and related to maps of the mass distribution in the Universe. The magnitude of the shear depends upon the combined effect of the gravitational potentials between the source and the observer. An observer will detect this integrated effect and maps of the integrated mass, or convergence, can be made. Gravitational lensing has the significant advantage that it is sensitive to both luminous and dark matter, and can therefore directly detect the combined matter distribution. In addition, weak lensing convergence maps allow for detecting the growth of structure in the Universe and hence they can also be used for probing statistics beyond two point correlation functions, such as in the higher moments of the convergence field or by observing the topology of the field with Minkowski functionals and peak statistics \citep{Dietrich2010, mawdsley2020}. As future surveys attempt to further probe the non-linear regime of structure growth, the information held in these higher order statistics will become increasingly important, and will also require accurate simulations in order to provide cosmological constraints. This requirement for large numbers of simulations that also model complex physical phenomena means that more computationally efficient alternatives to \textit{N}-body simulations, such as the GAN approach proposed in this work, are required.

In order to train the GAN algorithm to produce realistic convergence maps, we used publicly available datasets. In particular, to test whether we can reproduce the original results from \cite{Mustafa2019} we used the publicly available data from \cite{cosmogan}. The dataset consisted of 8000 weak lensing maps that were originally produced by running a Gadget2 \citep{volker2005} simulation with $512^{3}$ particles in $240$ $\mathrm{Mpc}/h$ box. To perform ray tracing the Gadget weak lensing simulation pipeline was used. The simulation box was rotated multiple times for each ray tracing procedure, resulting in 1000 12 sq. degree maps per simulation box. 

In order to train the GAN algorithm on convergence maps of different cosmologies and redshifts, we used the publicly available Columbia Lensing dataset described in \cite{matilla2016, gupta2018, columbia}. The available dataset contains weak lensing convergence maps covering a field of view of 3.5 deg $\times$ 3.5 deg, with resolution of 1024 $\times$ 1024 pixels. The maps were originally produced using Gadget2 DM-only simulation data with 240 Mpc$/h$ side cube and $512^{3}$ particles. The dataset includes 96 different cosmologies (with varying $\Omega_{m}$ and $\sigma_{8}$ parameters). The values of $\Omega_{m} = 0.260$ and $\sigma_{8} = 0.8$ were used as the fiducial cosmology. In our analysis, we only used a small subset of this dataset, namely, the maps where only one of the two cosmological parameter varies. In particular, we worked with the dataset consisting of the maps with $\sigma_{8} = \{0.436, 0.814 \}$ with a common value of $\Omega_{m} = 0.233$. This was done in order to simplify the latent space analysis.

For the weak lensing map data we used the same architecture as described in table 1 in \cite{rodriguez2018}. In fact the same basic architecture with minor variations was used for training all the datasets described later on (see tables \ref{generator_architecture} and \ref{discriminator_architecture}). In particular, for the cosmic web slice data we increased the random input vector size to 256 (from 64 in the case of weak lensing maps). For the multi-component dataset (dark matter, gas and internal energy), we changed the input layer to account for the 3-D input data. In summary, for the discriminator, we used an architecture of 4 convolutional layers with batch normalization and \textit{LeakyRelu} activation \citep{fukushima1980, Lecun1998, Nair2010, Maas2013, Ioffe2015}. In the final layer, we used the \textit{sigmoid} activation. For the generator, we used a linear input layer followed by 4 deconvolution layers with \textit{Relu} activation and batch normalization. The output layer used the $tanh$ activation function. The key parameter in terms of the training procedure is the learning rate. For all the cosmic web slice datasets, we found the learning rate value of $R_{L} = 3 \times 10^{-5}$ to work well. In the case of all the considered weak lensing datasets we used $R_{L} = 9 \times 10^{-6}$. The training procedure and all the key parameters are described in great detail in the publicly available code (see section \ref{data_code_availability} for more information). 

\subsection{Cosmic Web Slice Data}

The cosmic web or the dark matter overdensity field refers to the intricate network of filaments and voids as seen in the output data of \textit{N}-body simulations. The statistical features of the cosmic web contain important information about the underlying cosmology and could hide imprints of modifications to the standard laws of gravity. In addition, emulating a large number of overdensity fields is important for reliable estimation of the errors of cosmological parameters. Hence, emulators, such as the one proposed in this work, will be of special importance for the statistical analysis in the context of the upcoming observational surveys.      

To build the cosmic web training dataset we used a similar procedure to the one outlined in \cite{rodriguez2018}. In particular, we ran L-PICOLA \citep{howlett2015} to produce a total of 15 independent simulation boxes with different cosmologies. Initially, we used the same cosmology as described in \cite{rodriguez2018} with $h = 0.7$, $\Omega_{\Lambda} = 0.72$ and $\Omega_{m} = 0.28$. Subsequently, we studied the effects of varying one of the cosmological parameters, namely the $\sigma_{8}$ parameter. We explored the values of $\sigma_{8} = \{0.7,0.8,0.9\}$ along with $\Omega_{\Lambda} = 0.7$, $\Omega_{m} = 0.3$ and $h = 0.67$. For each different set of simulations, we saved snapshots at 3 different redshifts: $z = \{0.0, 0.5,1.0\}$. For each simulation, we used a box size of 512 Mpc/$h$ and a number of particles of $512^{3}$. For the latent space interpolation procedure, we trained the GAN on slices with redshifts $\{0.0,1.0\}$, with a common value of $\sigma_{8} = 0.8$.

To produce the slices for training the GAN, we used \textit{nbodykit} \citep{nbodykit}, which allows painting an overdensity field from a catalogue of simulated particles. To obtain the needed slices, we cut the simulation box into sections of 2 Mpc width in $x,y,z$ directions and for each section a mesh painting procedure was done. This refers to splitting the section into cells, where the numerical value of each cell corresponds to the dark matter overdensity $1 + \delta(x)$. Finally, after a 2-D projection of each slice, a $256^{2}$ px image was obtained, with each pixel value corresponding to the overdensity field. To emphasize the features of the large scale structure, we applied the same non-linear transformation as described in \cite{rodriguez2018}: $s(x) = 2x/(x+a) - 1$, with $a = 250$, which rescales the overdensity values to $[-1,1]$ and increases the contrast of the images.

In order to emulate modified gravity effects we used the MG-PICOLA code, which extends the original L-PICOLA code in order to allow simulating theories that exhibit scale-dependent growth \citep{scoccimarro2012, Tassev_2013, Winther_2017, mg-picola}. This includes models, such as $f(R)$ theories which replace the Ricci scalar with a more general function in the Einstein-Hilbert action (see \cite{Li2019B} for an overview of the phenomenology of such models). In particular, multiple runs of MG-PICOLA were run with the following range of the $f_{R0}$ parameter: $[10^{-7}, 10^{-1}]$. Such a wide range was chosen to make the latent space interpolation procedure easier. The $f(R)$ simulations were also run with the same seed as the corresponding $\Lambda$CDM simulations, making the two datasets described above directly comparable.

\label{cosmic-web data}

\subsection{Dark Matter, Gas and Internal Energy Data}

Simultaneously generating the dark matter and the corresponding baryonic overdensity field data is a great challenge from both the theoretical and the computational perspectives. Namely, generating the baryonic distribution requires detailed hydrodynamical simulations that account for the intricacies of galaxy formation and feedback processes, which leads to a major increase in the required computational resources. For this reason, emulating large amounts of hydrodynamical simulation data is of special importance.   

To produce the dark matter, baryonic matter and the internal energy distribution slices we used the publicly available Illustris-3 simulation data \citep{vogelsberger2014,nelson2015}. Illustris-3 refers to the low resolution Illustris run including the full physics model with a box size of 75000 $\textrm{kpc}/h$ and over $9 \times 10^{7}$ dark matter and gas tracer particles. The cosmology of the simulation can be summarized by the following parameters: $\Omega_{m} = 0.2726$, $\Omega_{\Lambda} = 0.7274$, $h =0.704$. The simulation included the following physical effects: radiative gas cooling, star formation, galactic-scale winds from star formation feedback, supermassive blackhole formation, accretion, and feedback. 

\begin{figure*}
\centering
\captionsetup[subfigure]{justification=centering}
  \begin{subfigure}[b]{0.34\textwidth}
  \centering
    \includegraphics[width=0.8\textwidth]{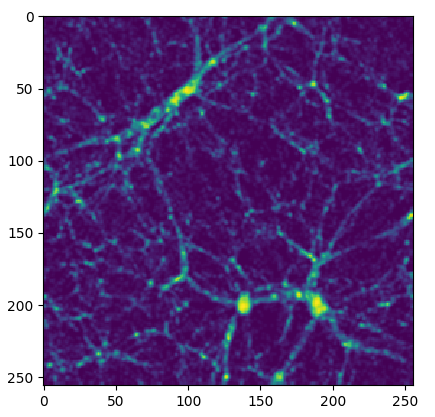}
    \caption{DM overdensity field}
    \label{fig3:a1}
    
  \end{subfigure}
  \begin{subfigure}[b]{0.334\textwidth}
  \centering
    \includegraphics[width=0.8\textwidth]{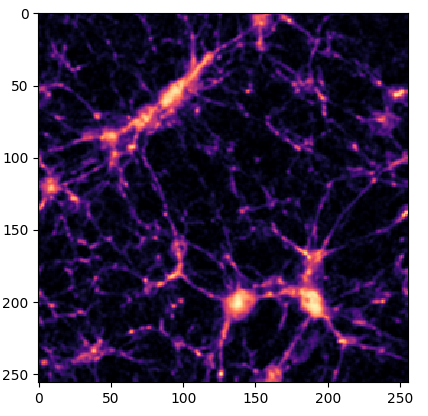}
    \caption{Gas overdensity field}
    \label{fig3:a2}
  \end{subfigure}
    \begin{subfigure}[b]{0.34\textwidth}
  \centering
    \includegraphics[width=0.8\textwidth]{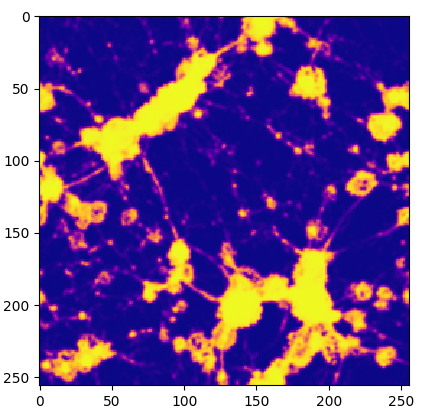}
    \caption{Internal energy field}
    \label{fi3:a3}
  \end{subfigure}
    \begin{subfigure}[b]{0.34\textwidth}
  \centering
    \includegraphics[width=0.8\textwidth]{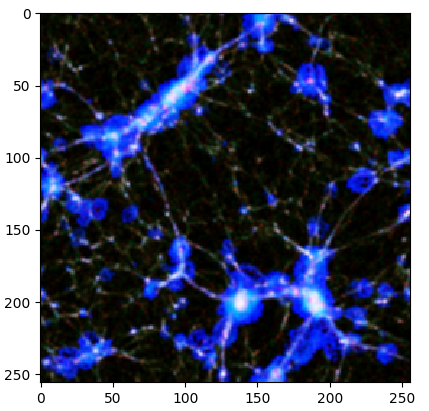}
    \caption{All components combined}
    \label{fig3:a4}
  \end{subfigure}
  \caption{Samples from the Illustris simulation dataset used to train the GAN algorithm: 2-D slices of the different simulation components.}
  \label{figure3}
\end{figure*}

To form the training dataset we used an analogous procedure to the one used for the cosmic web slices in section \ref{cosmic-web data}. In particular, we sliced the full simulation box into slices of 100 $\textrm{kpc}/h$ and for each slice used mesh painting to obtain an overdensity field. This was done for the dark matter and gas data. In addition, we also used the available internal energy (thermal energy in the units of $(\rm km/s)^{2}$) distribution data. Fig. \ref{figure3} shows a few samples from the dataset.

To investigate whether the GAN algorithm could be trained on multidimensional array data, we treated the DM, gas and energy distribution 2-D slices as RGB planes in a single image. In particular, a common way of representing colors in an image is forming a full color image out of three planes, each corresponding to the pixel values for red, green and blue colours. In this framework, a full-color image corresponds to a 3-D array. Convolutional neural networks, including the one that the \textit{cosmoGAN} algorithm is based on are originally designed to be trained on such RGB images. Hence we combined the mentioned DM, gas and internal energy slices into a set of RGB arrays that were used as a training set.  

\subsection{The Training Procedure}

The initial stages of training (i.e. reproducing the results in \cite{rodriguez2018, Mustafa2019}) were done using the Google Cloud computing platform. The following setup was used: 4 standard vCPUs with 15 GB memory, 1 NVIDIA Tesla K80 GPU and 2TB of SSD hard drive space. We found the typical training time using this setup to be around 48 hours.  This corresponds to an estimated CO$_{2}$ emission of 8.93 kg CO$_2$ eq. of which 100 percent were directly offset by the cloud provider \citep{Lacoste2019}. 

The later stages of training (i.e. training the GAN on different cosmology, modified gravity and redshift data) were done using the local Sciama HPC cluster, which has 3702 cores of 2.66 GHz Intel Xeon processors with 2 GB of memory per core. The training procedure was found to be at least 5 times slower when compared to the outlined Google Cloud setup, as the HPC did not have access to a functional GPU and an older version of Tensorflow as used.

It is important to note that the GAN training procedure is significantly longer than running a small conventional simulation (e.g. a few hours per L-PICOLA simulation). However the advantage of GANs lies in the fact that once the training is done, thousands of novel samples can be produced in a matter of seconds.

Given how unstable the GAN training procedure is we used a simple procedure of evaluating the best checkpoint: we calculated the mean square difference between the mean values of the GAN-produced and the test dataset power spectra, pixel histograms and the Minkowski functionals. The set of GAN weights that minimize this value was used for the plots displayed in the result section.  

\section{Diagnostics}

The results produced by the algorithm were investigated using the following diagnostics: the 2-D matter power spectrum, overdensity (pixel) value histogram and the three Minkowski functionals. In addition, we computed the cross and the auto power spectrum in order to investigate the correlations between the datasets on different scales. The cross power spectrum was calculated using: 

\begin{equation}
    \langle \Bar{\delta_{1}}(l)\Bar{\delta_{2}^{*}}(l^{'}) \rangle = (2 \pi)^{2}\delta_{D}(l-l^{'})P_{\times}(l),
\end{equation}

\noindent where $\Bar{\delta_{1}}$ and  $\Bar{\delta_{2}^{*}}$ are the Fourier transforms of the two overdensity fields at some Fourier bin $l$ and $\delta_{D}$ is the Dirac delta function.

The Minkowski functionals are a useful tool in studying the morphological features of fields that provide not only the information of spatial correlations but also the information on object shapes and topology. For some field $f(x)$ in 2-D space we can define the three Minkowski functionals as follows: 

\begin{equation}
    V_{0}(\nu) = \int^{}_{Q_{\nu}} d \Omega,
    \quad\text{}\quad
    V_{1}(\nu) = \int^{}_{\partial Q_{\nu}} \frac{1}{4}dl,
    \quad\text{}\quad
    V_{2}(\nu) = \int^{}_{\partial Q_{\nu}} \frac{\kappa_{c}dl}{2\pi} .
    \label{minkowski}
\end{equation}

Where $Q_{\nu} \equiv \{x \in \R^{2} | f(x) > \nu \} $ is the area and $\partial Q_{\nu} \equiv \{x \in \R^{2} | f(x) = \nu \} $ is the boundary of the field above threshold value $\nu$. The integrals $V_{0}$, $V_{1}$, $V_{2}$ correspond to the area, boundary length and the integrated geodesic curvature $\kappa_{c}$ along the boundary. In simple words, the procedure of measuring the Minkowski functionals refers to taking the values of the field at and above a given threshold $\nu$, evaluating the integrals in eq. \ref{minkowski} and then changing the threshold for a range of values. 

Minkowski fuctionals are a useful tool in weak lensing convergence map studies as they allow us to capture non-Gaussian information on the small scales, which is not fully accessed by the power spectrum alone. In addition, Minkowski functionals have been used to detect different cosmologies, modified gravity models and the effects of massive neutrinos in weak lensing convergence maps \citep{petri2013, Ling2015L, marques2019}. Given the usefulness of Minkowski functionals in accessing the non-Gaussian information on the small scales, we chose to apply the functionals for studying the produced cosmic web projections as well. To calculate the Minkowski functionals properly on a 2-D grid we used the \textit{minkfncts2d} algorithm, which utilizes a marching square algorithm as well as pixel weighting to capture the boundary lengths correctly \citep{mantz2008,minkfncts2d}. 

Minkowski functionals are sensitive to the Gaussian smoothing applied to the GAN-produced images and the training data. Hence, it is important to study the effects of Gaussian smoothing as it might give a deeper insight into the detected differences between the datasets. The procedure of smoothing refers to a convolution between a chosen kernel and the pixels of an image. In more detail, a chosen kernel matrix is centered on each pixel of an image and each surounding pixel is multiplied by the values of the kernel and subsequently summed. In the simplest case, such a procedure corresponds to averaging a chosen number of  pixels in a given image. In the case of Gaussian filtering, a Gaussian kernel is used instead.   

\begin{figure*}
\centering
\captionsetup[subfigure]{justification=centering}
  \begin{subfigure}[b]{0.27\textwidth}
    \includegraphics[width=\textwidth]{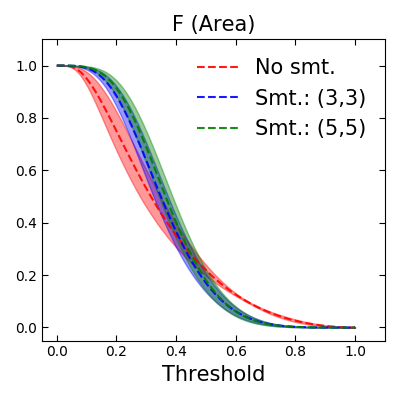}
    \label{MF_smoothing:1}
    
  \end{subfigure}
  \begin{subfigure}[b]{0.27\textwidth}
    \includegraphics[width=\textwidth]{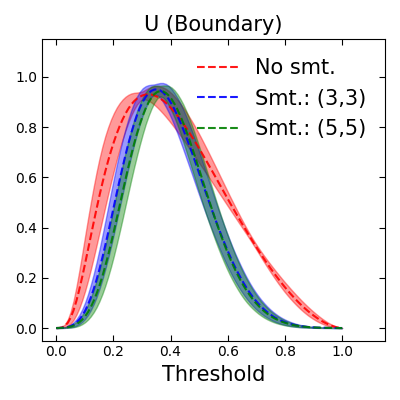}
    \label{MF_smoothing:2}
  \end{subfigure}
    \begin{subfigure}[b]{0.27\textwidth}
    \includegraphics[width=\textwidth]{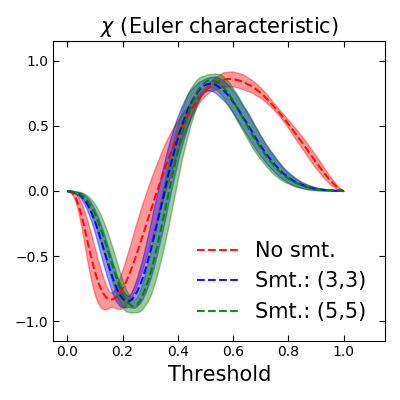}
    \label{MF_smoothing:3}
  \end{subfigure}
  \caption{An illustration of the effects of Gaussian smoothing on the Minkowski functionals calculated using cosmic web slices from the training data with redshift $z = 0.0$. The colored bands correspond to the mean and the standard deviation of the functionals calculated using different sizes of Gaussian smoothing kernels on a batch of 64 images. Smt. refers to Gaussian smoothing with the numbers in the brackets corresponding to the kernel size.}
  \label{MF_smoothing_effects}
\end{figure*}

To filter the noise we used Gaussian smoothing with a $3 \times 3$ kernel window and a standard deviation of 1 px. We found the Minkowski functionals to be especially sensitive to any kind of smoothing. For instance, the position and the shape of the trough of the third Minkowski functional is highly sensitive to existence of any small-scale noise. Fig. \ref{MF_smoothing_effects} illustrates the effects of Gaussian smoothing with different kernel sizes on the three Minkowski functionals. 

In addition to the outlined diagnostics, we also tested the uniqueness of the GAN-produced samples. More specifically, in order to show that the GAN is indeed producing novel data rather than just mimicking the training dataset, we compared the GAN-produced samples with the training samples. A 1000 GAN-generated samples were compared against each sample in the training dataset by calculating the difference in the power spectra, pixel intensity histogram and the Minkowski functionals. For all the relevant datasets it was found that the GAN algorithm was indeed producing novel data rather than just mimicking the training data, which aggrees well with the previous results found in \cite{Mustafa2019}.

\section{Results}

\subsection{Weak Lensing Map Results}
\label{cosmoGAN_original_results}

After around 150 epochs (corresponding to around 120 hours on our HPC) the GAN started producing statistically realistic convergence maps as measured by the power spectrum and the Minkowski functionals. The diagnostics were computed at an ensemble level -- 100 batches of 64 convergence maps were produced by the GAN and the mean values along with the standard deviation were computed and compared against the training and the test (simulation data not used in the training procedure) datasets. 
An analogous procedure was done when calculating the pixel intensity distribution histograms. 

The two mentioned datasets were obtained by splitting the total available simulation data into $2/3$ dedicated to the training dataset, while the other $1/3$ of the data was used as the test dataset. The same general setup was used when testing all the other datasets. 

\begin{figure*}
\centering
\captionsetup[subfigure]{justification=centering}
  \begin{subfigure}[b]{0.48\textwidth}
  \centering
    \includegraphics[width=0.8\textwidth]{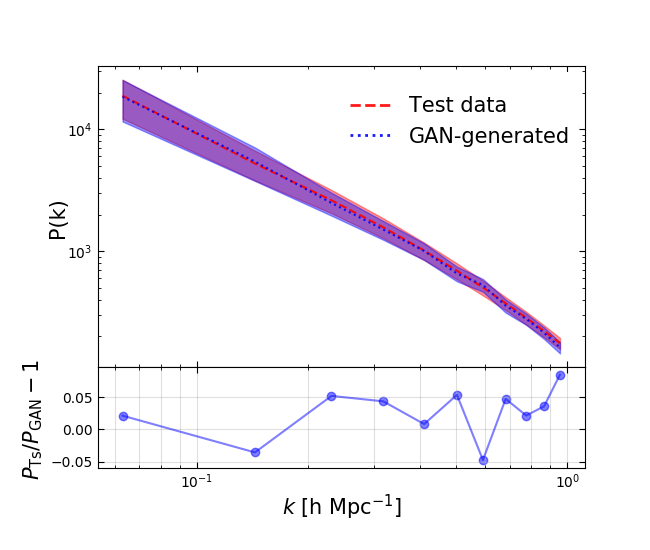}
    \caption{Power spectrum}
    \label{fig4:a1}
    
  \end{subfigure}
  \begin{subfigure}[b]{0.48\textwidth}
  \centering
    \includegraphics[width=0.8\textwidth]{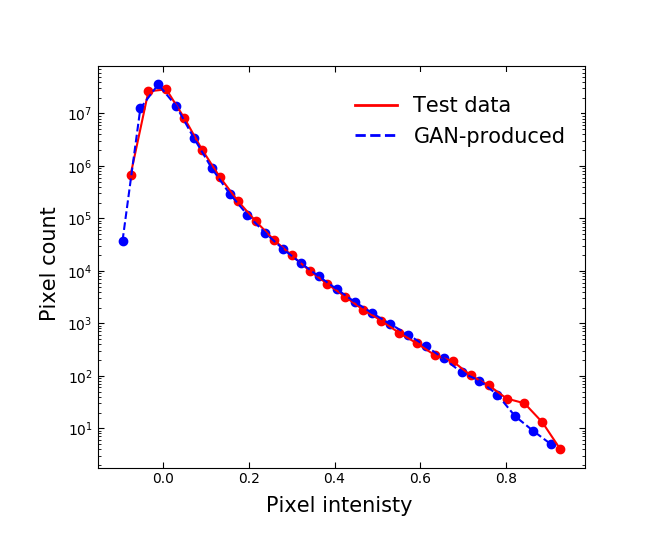}
    \caption{Pixel intensity histogram}
    \label{fig4:a2}
  \end{subfigure}
  \caption{The matter power spectrum (with the relative difference) and the pixel intensity histogram for an ensemble of 6400 weak lensing convergence maps. The dashed lines correspond to the mean values, while the contours correspond to the standard deviation. Note that the pixel intensity values were normalized to the range of $[-1,1]$. }
  \label{figure4}
\end{figure*}

The power spectra agree well between the GAN-produced and the training and test data, with minor differences on the small scales (see fig. \ref{figure4}). In particular, the difference between the test and the GAN-produced dataset power spectra is around 5\% or lower for most values of $k$. Only at the smallest scales, a significant difference of 10\% is reached. Similarly, the pixel intensity histogram in general shows a good agreement with significant differences appearing only for the highest and the lowest pixel intensity values (which is also detected in the original work in \cite{Mustafa2019}). A selection of GAN-produced maps are presented for visual inspection in fig. \ref{cosmoGAN_samples}.

\begin{figure*}
\centering
\captionsetup[subfigure]{justification=centering}
  \begin{subfigure}[b]{0.325\textwidth}
  \centering
    \includegraphics[width=0.8\textwidth]{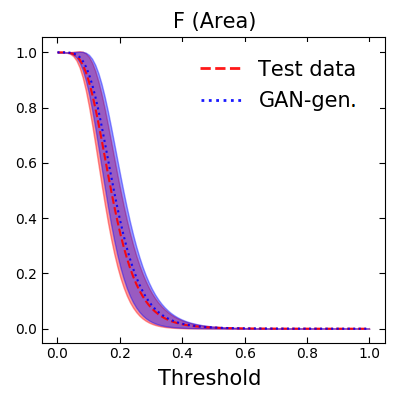}
    \label{fig5:a1}
    
  \end{subfigure}
  \begin{subfigure}[b]{0.325\textwidth}
  \centering
    \includegraphics[width=0.8\textwidth]{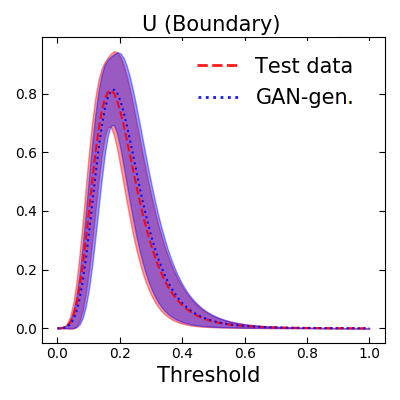}
    \label{fig5:a2}
  \end{subfigure}
    \begin{subfigure}[b]{0.325\textwidth}
  \centering
    \includegraphics[width=0.8\textwidth]{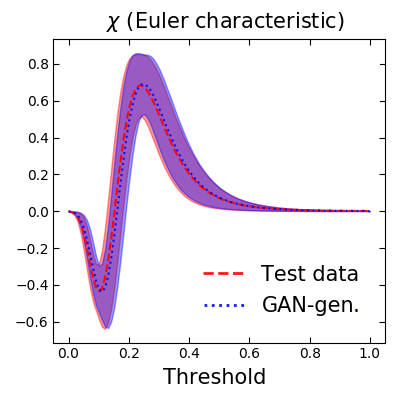}
    \label{fig5:a3}
  \end{subfigure}
  \caption{A comparison of the Minkowski functionals evaluated using 100 batches of 64 randomly selected maps for both datasets.}
  \label{figure5}
\end{figure*}

We also computed the Minkowski functionals for the GAN-produced and the test datasets. The results are shown in fig. \ref{figure5}. In general there is a good agreement between the test data and the GAN-produced maps, given the standard deviation, however, some minor differences can be detected in the Euler characteristic and the boundary functional, likely resulting from noise. 

\subsection{Weak Lensing Maps of Multiple Cosmologies}
\label{section_weak_lensing_multiple_cosmologies}

We also found that the GAN is capable of producing realistic weak lensing maps for multiple cosmologies. This is an important result as it shows that the algorithm is able to pick up on the various subtle statistical differences between different cosmologies that usually requires a detailed study of the power spectrum, Minkowski functionals and other statistics.
 
However, we found the training procedure to be highly prone to mode collapse. A wide hyperparameter search had to be performed to find an optimal set of parameters that did not lead to full or partial mode collapse. The most important parameter in this context was found to be the learning rate. As a rule of thumb, decreasing the learning rate led to mode collapse happening later in the training procedure. When the learning rate was reduced below a certain value (discussed further in the analysis section), mode collapse was avoided altogether. As in the case with the cosmic web slice data, applying a transformation to each pixel of the image in order to increase the contrast had a positive effect in reducing the probability of mode collapse as well.

 \begin{figure*}
\centering
\captionsetup[subfigure]{justification=centering}
  \begin{subfigure}[b]{0.48\textwidth}
  \centering
    \includegraphics[width=0.8\textwidth]{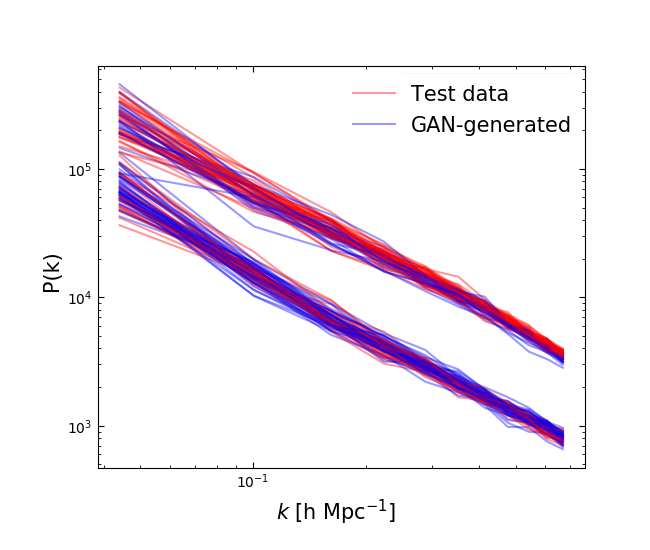}
    \label{shear_maps_sigma:1}
    
  \end{subfigure}
  \begin{subfigure}[b]{0.48\textwidth}
  \centering
    \includegraphics[width=0.8\textwidth]{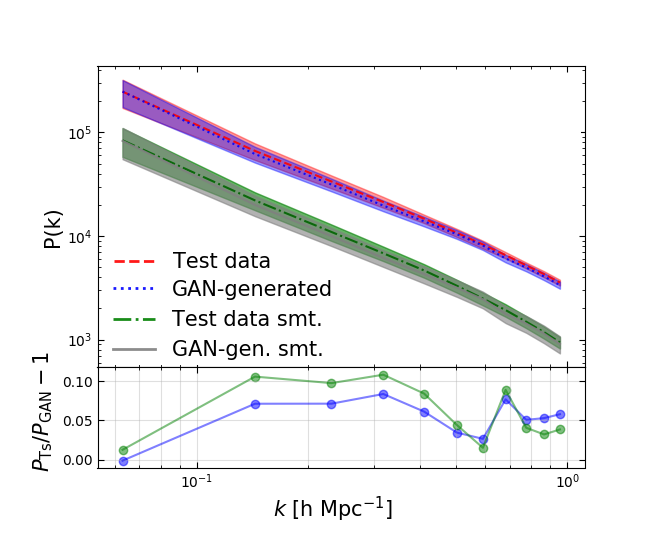}
    \label{shear_maps_sigma:2}
  \end{subfigure}
    \begin{subfigure}[b]{0.48\textwidth}
  \centering
    \includegraphics[width=0.8\textwidth]{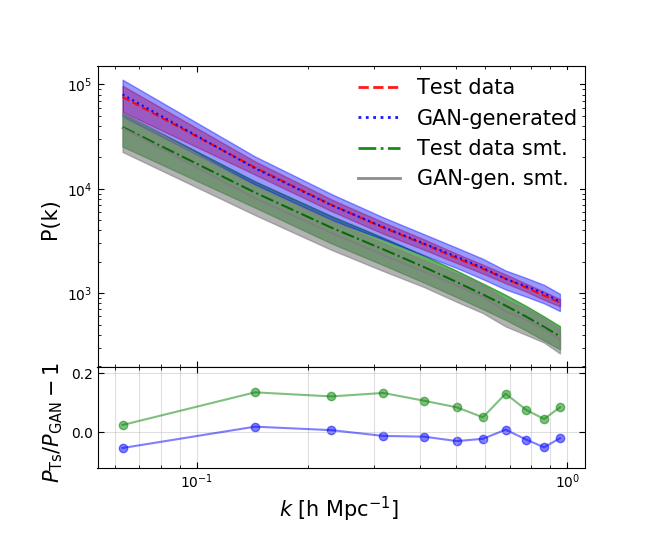}
    \label{shear_maps_sigma:3}
  \end{subfigure}
    \begin{subfigure}[b]{0.48\textwidth}
  \centering
    \includegraphics[width=0.8\textwidth]{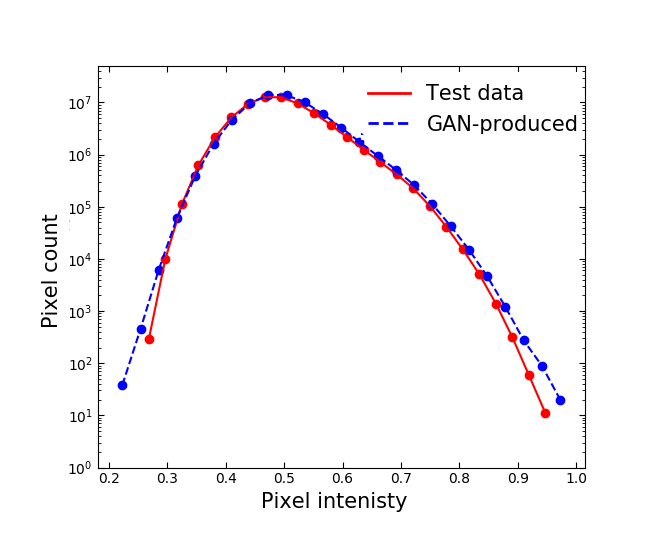}
    \label{shear_maps_sigma:4}
  \end{subfigure}
  \caption{A selection of diagnostics to compare the test and the GAN-produced weak lensing convergence maps for $\sigma_{8}$ = $\{0.436,0.814\}$ with $\Omega_{m} = 0.233$. \textbf{Top left}: power spectra for an ensemble of 64 randomly chosen shear maps; \textbf{top right}: power spectra (mean and standard deviation) with and without Gaussian smoothing produced using 1000 randomly chosen shear maps with $\sigma_{8} = 0.814$; \textbf{bottom left}: same as top right, but for $\sigma_{8} = 0.436$; \textbf{bottom right}: the pixel intensity distribution (for both datasets combined). The blue and the green dots give  $P_{\rm Ts}/P_{\rm GAN} - 1$ with and without Gaussian smoothing applied correspondingly.   }
  \label{shear_maps_sigma}
\end{figure*}

Fig. \ref{shear_maps_sigma} summarizes the results of training the GAN on shear maps with different $\sigma_{8}$ values. The results indicate an agreement of the power spectra in the range of 5-10\% for $k > 10^{-1}$ $\mathrm{h}$ $\mathrm{Mpc^{-1}}$ for $\sigma_{8} = 0.814$. In the case of $\sigma_{8} = 0.436$ the agreement is significantly better, ranging between 1-3\% on most scales. Interestingly, Gaussian smoothing  increases the difference to around 5-15\% in this particular case. This shows that for this dataset, Gaussian noise is not the major source of the statistical differences between the training and the GAN-generated datasets.   

\begin{figure*}
\centering
\captionsetup[subfigure]{justification=centering}
  \begin{subfigure}[b]{0.325\textwidth}
  \centering
    \includegraphics[width=0.8\textwidth]{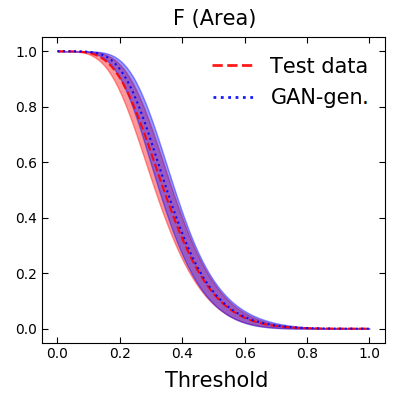}
    \label{shear_map_sigma_MFs:1}
    
  \end{subfigure}
  \begin{subfigure}[b]{0.325\textwidth}
  \centering
    \includegraphics[width=0.8\textwidth]{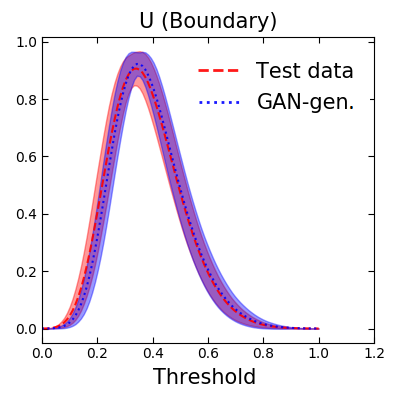}
    \label{shear_map_sigma_MFs:2}
  \end{subfigure}
    \begin{subfigure}[b]{0.325\textwidth}
  \centering
    \includegraphics[width=0.8\textwidth]{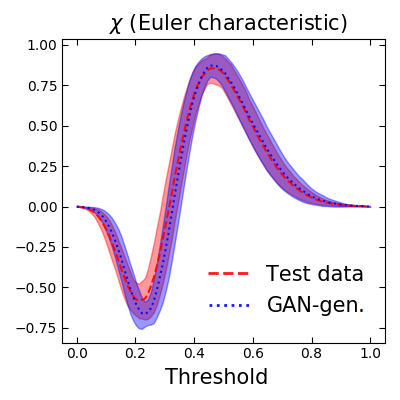}
    \label{shear_map_sigma_MFs:3}
  \end{subfigure}
  \caption{ A comparison of the Minkowski functionals evaluated using 1000 randomly selected weak lensing convergence maps with $\sigma_{8} = \{0.436,0.814\}$. Gaussian smoothing is applied for all datasets. }
  \label{shear_map_sigma_MFs}
\end{figure*}

Fig. \ref{shear_map_sigma_MFs} compares the Minkowski functionals calculated using the training and the GAN-produced datasets. Given the standard deviation in both datasets, the results overlap for all threshold values. However, for thresholds in the range of $[0.0,0.4]$ there is a significant difference between the training and the GAN-generated datasets. We found that this is partially due to small-scale noise in the GAN-produced data (see fig. \ref{MF_smoothing_effects}). However, after experimenting with adding artificial noise to the training dataset images, it is clear that the noise alone cannot fully account for the observed differences in the Minkowski functionals. Another reason for the observed differences could be a relatively small size of the used dataset consisting of a few thousand weak lensing maps. It is likely that having more training data samples could significantly improve the results.    

\subsection{Cosmic Web for Multiple Redshifts}
\label{section_cosmic_web_redshifts}

\begin{figure*}
\centering
\captionsetup[subfigure]{justification=centering}
  \begin{subfigure}[b]{0.48\textwidth}
    \centering
    \includegraphics[width=0.8\textwidth]{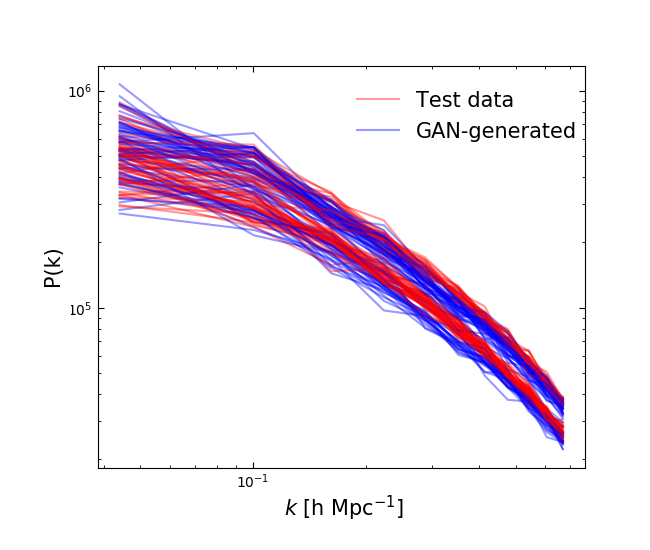}
    \label{fig6:a1}
  \end{subfigure}
  \begin{subfigure}[b]{0.48\textwidth}
    \centering
    \includegraphics[width=0.8\textwidth]{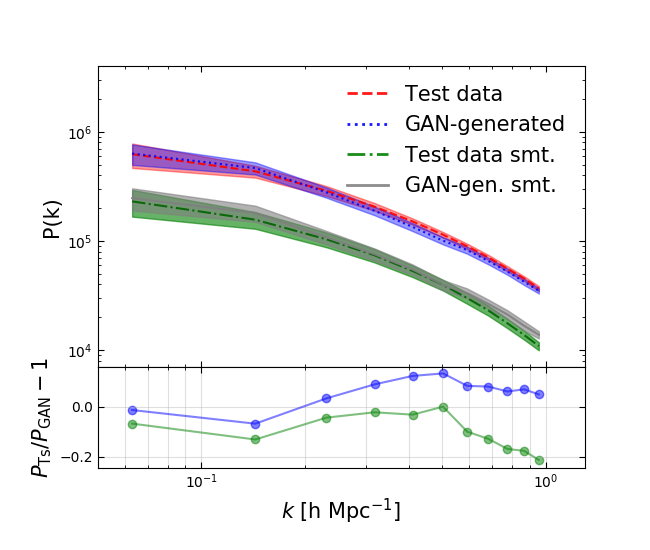}
    \label{fig6:a2}
  \end{subfigure}
    \begin{subfigure}[b]{0.48\textwidth}
      \centering
    \includegraphics[width=0.8\textwidth]{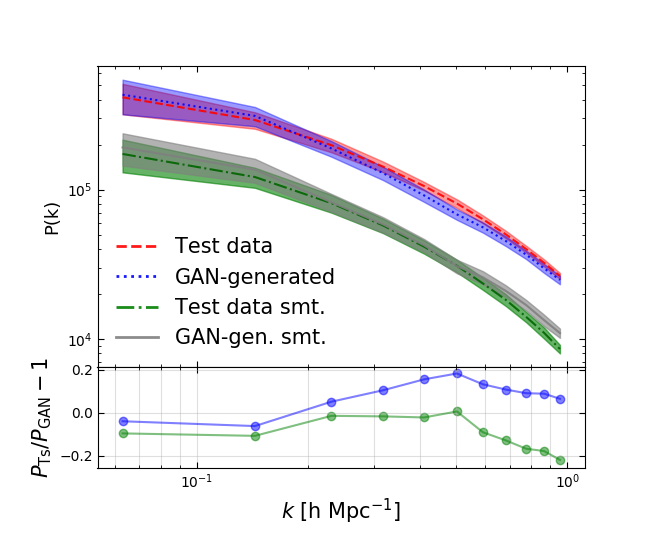}
    \label{fig6:a3}
  \end{subfigure}
    \begin{subfigure}[b]{0.48\textwidth}
      \centering
    \includegraphics[width=0.8\textwidth]{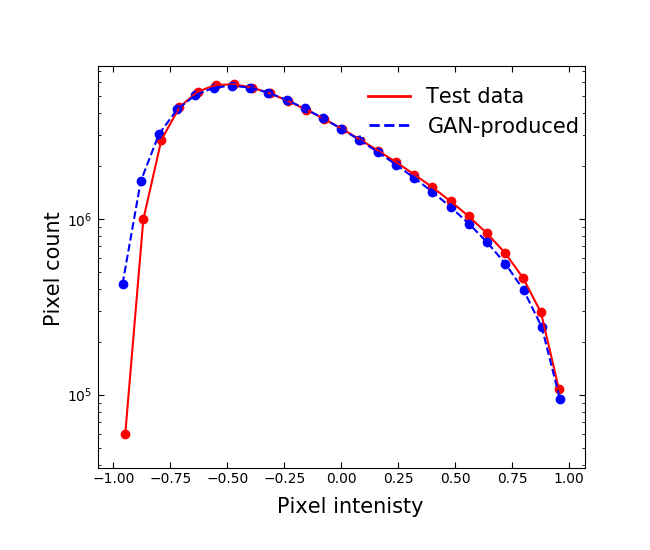}
    \label{fig6:a4}
  \end{subfigure}
  \caption{A selection of diagnostics to compare the test and the GAN-produced cosmic web slices for redshifts $z=0.0$ and $z=1.0$ with $\sigma_{8}= 0.8$. \textbf{Top left}: power spectra for an ensemble of 64 randomly chosen slices for two different redshifts; \textbf{top right}: mean and standard deviation of the power spectra produced using 1000 randomly chosen slices with $z=0.0$; \textbf{bottom left}: same as top right, but for $z=1.0$; \textbf{bottom right}: the overdensity histogram (no smoothing). The blue and the green dots give $P_{\rm Ts}/P_{\rm GAN} - 1$ with and without Gaussian smoothing applied correspondingly. }
  \label{figure6}
\end{figure*}

\begin{figure*}
\centering
\captionsetup[subfigure]{justification=centering}
  \begin{subfigure}[b]{0.325\textwidth}
  \centering
    \includegraphics[width=0.8\textwidth]{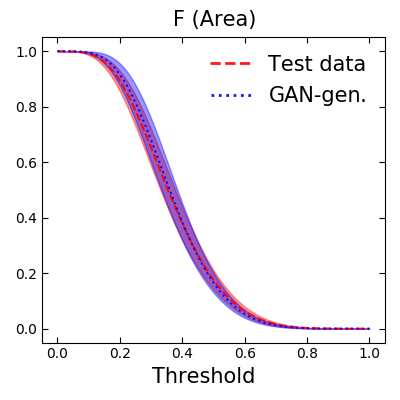}
    \label{fig7:a1}
    
  \end{subfigure}
  \begin{subfigure}[b]{0.325\textwidth}
  \centering
    \includegraphics[width=0.8\textwidth]{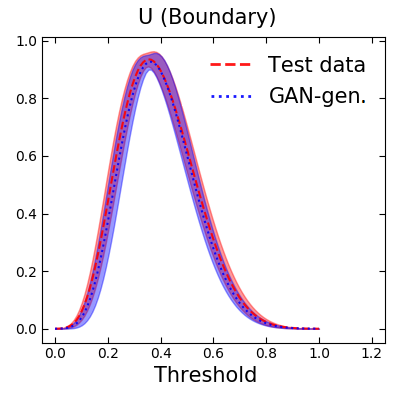}
    \label{fig7:a2}
  \end{subfigure}
    \begin{subfigure}[b]{0.325\textwidth}
  \centering
    \includegraphics[width=0.8\textwidth]{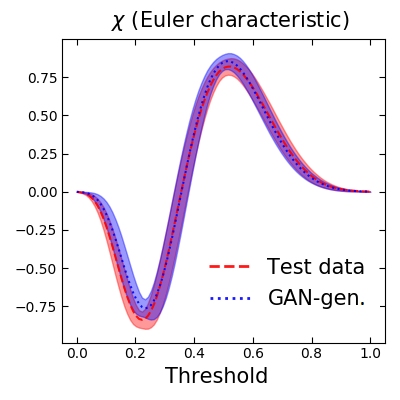}
    \label{fig7:a3}
  \end{subfigure}
  \caption{A comparison of the Minkowski functionals evaluated using 1000 randomly selected cosmic web slices of redshifts $z = \{0.0,1.0\}$ for both datasets. Gaussian smoothing is applied for all datasets. }
  \label{figure7}
\end{figure*}

We also found that the GAN is capable of producing realistic cosmic web 2-D projections for different redshifts. As before with the weak lensing maps of different cosmologies, this result illustrates that the algorithm in general does not get \textit{confused} between the two different redshifts and is capable of detecting subtle statistical differences between the different datasets (fig. \ref{figure6}). In addition, we found that using Gaussian smoothing, as before, led to a better agreement between the training and the GAN-produced datasets. The effect is especially noticeable in the Minkowski functional analysis (fig. \ref{figure7}). Visual samples of the produced cosmic web slices are shown in fig. \ref{cosmoGAN_cw_samples}.

We found the power spectra results for both redshift values to be very similar. Namely, for the non-smoothed case the difference between the test and the GAN-produced power spectra ranges between 5-10\%. The results are similar for the smoothed case, with exception of $k$ values around 1 $\mathrm{h}$ $\mathrm{Mpc^{-1}}$ where the difference reaches 20\%. 

The effects of the Gaussian smoothing on both the power spectra and the Minkowski functionals illustrate that one of the reasons for the differences between the GAN-generated and the test datasets is noise appearing on different scales in the GAN-produced images. Applying Gaussian smoothing, in general, filters the majority of such noise, however, it cannot fully account for all the differences appearing in the different statistical diagnostics. In addition, smoothing can improve the results on some scales, while worsening them on others. As an example, in fig. \ref{figure6}, Gaussian smoothing increases the difference between the GAN-produced and the test dataset power spectra on the smallest scales.

\subsection{Cosmic Web for Multiple Cosmologies and Modified Gravity Models}

Training the GAN on the cosmic web slices of different cosmologies and modified gravity models offered another way of testing whether the algorithm would pick up on the subtle statistical differences between the different datasets. In addition, the classification task for the discriminator neural network is more difficult when training on datasets with multiple cosmologies leading to longer training times.  

\begin{figure*}
\centering
\captionsetup[subfigure]{justification=centering}
  \begin{subfigure}[b]{0.48\textwidth}
  \centering
    \includegraphics[width=0.8\textwidth]{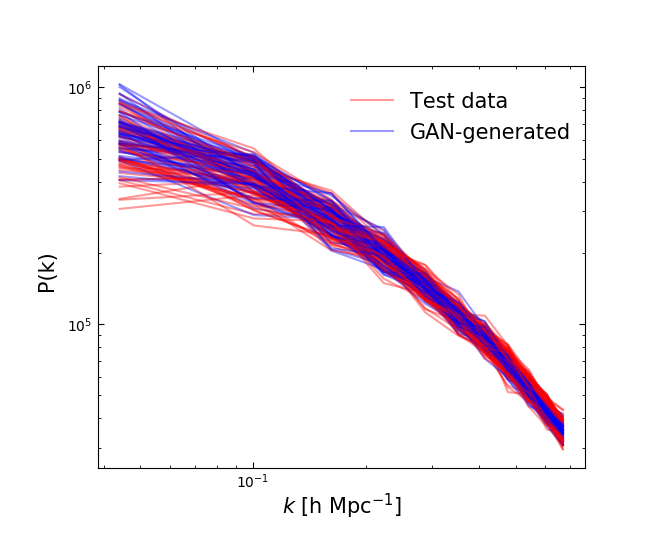}
    \label{fig8:a1}
  \end{subfigure}
  \begin{subfigure}[b]{0.48\textwidth}
  \centering
    \includegraphics[width=0.8\textwidth]{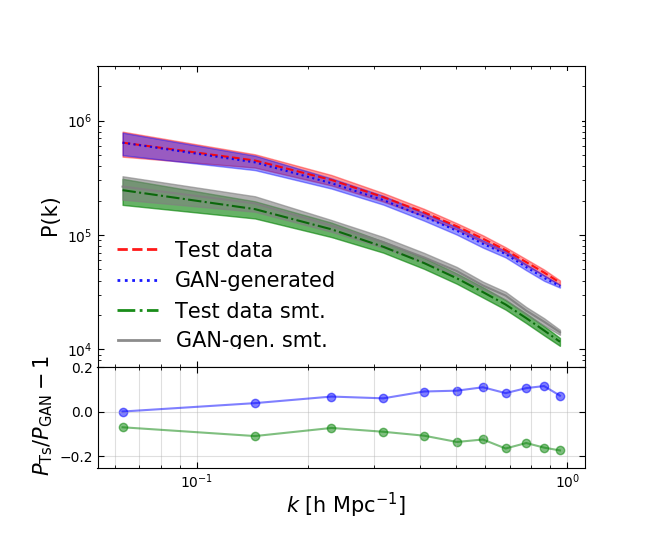}
    \label{fig8:a2}
  \end{subfigure}
    \begin{subfigure}[b]{0.48\textwidth}
  \centering
    \includegraphics[width=0.8\textwidth]{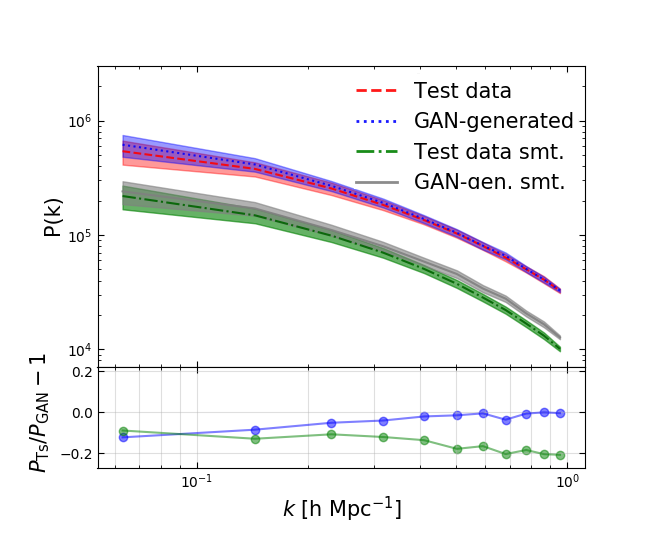}
    \label{fig8:a3}
  \end{subfigure}
    \begin{subfigure}[b]{0.48\textwidth}
  \centering
    \includegraphics[width=0.8\textwidth]{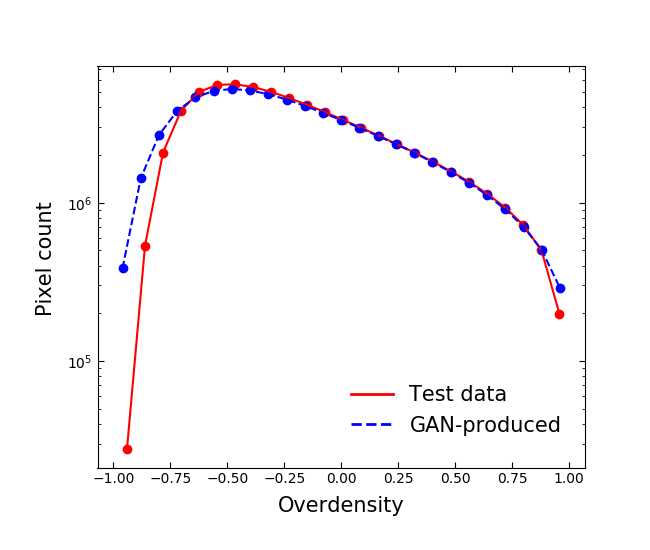}
    \label{fig8:a4}
  \end{subfigure}
  \caption{A selection of diagnostics to compare the test and the GAN-produced cosmic web slices for $\sigma_{8} = 0.7$ and $\sigma_{8} = 0.9$ at $z = 0.0$. \textbf{Top left}: power spectra for an ensemble of 64 randomly chosen slices for both datasets; \textbf{top right}: mean and standard deviation of the power spectra computed using 1000 randomly chosen slices of $\sigma_{8} = 0.9$; \textbf{bottom left}: same as top right, but for $\sigma_{8} = 0.7$; \textbf{bottom right}: the overdensity histogram (no smoothing). The blue and the green dots give  $P_{\rm Ts}/P_{\rm GAN} - 1$ with and without Gaussian smoothing applied correspondingly. }
  \label{figure8}
\end{figure*}

The results indicate that the GAN is indeed capable of producing statistically realistic cosmic web data of different cosmologies and modified gravity models. With no Gaussian smoothing applied, the relative agreement between the power spectra is 1-10\% (see fig. \ref{figure8}). Applying smoothing in this case resulted in increasing the relative power spectrum difference to over 10\% on average. In the case of cosmic web slices for different $f_{R0}$ values, the agreement between the two datasets was good, ranging between 1-10\% on all scales. Smoothing improved the situation only in the mid-range of the covered $k$ values, reducing the agreement on the smallest scales (see fig. \ref{cw_fR_plots}). 

\begin{figure*}
\centering
\captionsetup[subfigure]{justification=centering}
  \begin{subfigure}[b]{0.325\textwidth}
  \centering
    \includegraphics[width=0.8\textwidth]{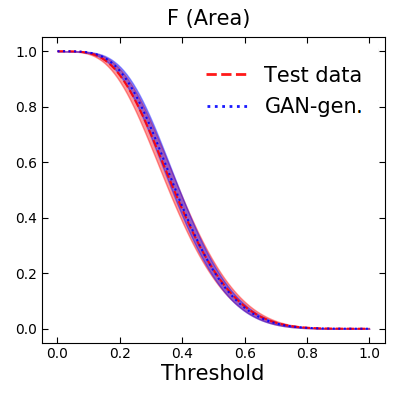}
    \label{fig9:a1}
    
  \end{subfigure}
  \begin{subfigure}[b]{0.325\textwidth}
  \centering
    \includegraphics[width=0.8\textwidth]{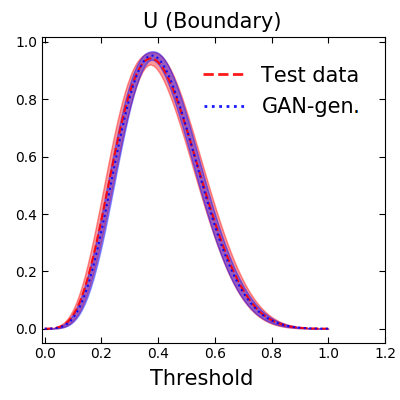}
    \label{fig9:a2}
  \end{subfigure}
    \begin{subfigure}[b]{0.325\textwidth}
  \centering
    \includegraphics[width=0.8\textwidth]{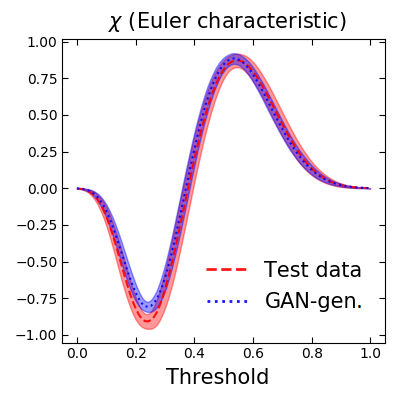}
    \label{fig9:a3}
  \end{subfigure}
  \caption{A comparison of the Minkowski functionals evaluated using 1000 randomly selected cosmic web slices from the dataset with two different values of $\sigma_{8} = \{0.7,0.9\}$. Gaussian smoothing is applied for both datasets. }
  \label{figure9}
\end{figure*}

\begin{figure*}
\centering
\captionsetup[subfigure]{justification=centering}
  \begin{subfigure}[b]{0.45\textwidth}
  \centering
    \includegraphics[width=0.8\textwidth]{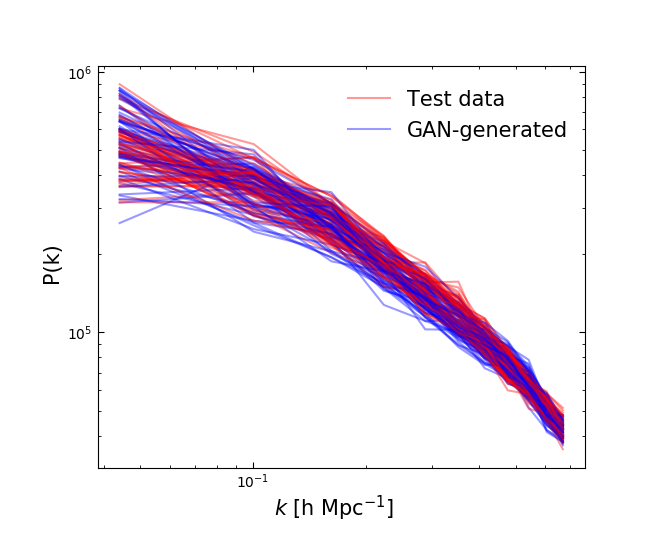}
    \label{cw_fR_plots:1}
  \end{subfigure}
  \begin{subfigure}[b]{0.45\textwidth}
  \centering
    \includegraphics[width=0.8\textwidth]{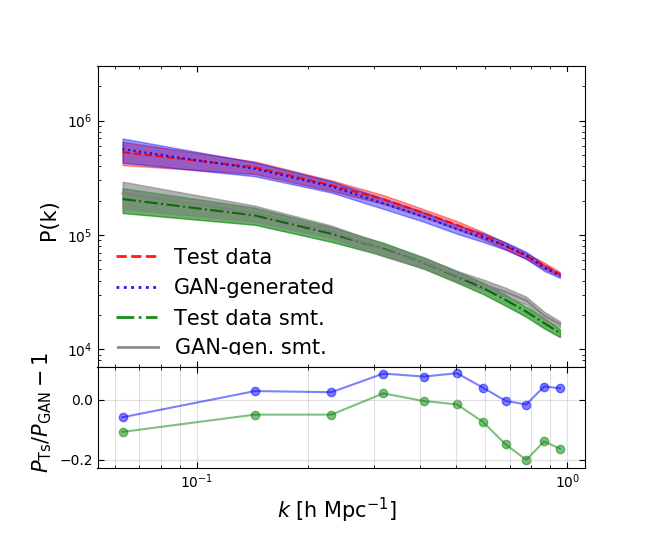}
    \label{cw_fR_plots:2}
  \end{subfigure}
    \begin{subfigure}[b]{0.45\textwidth}
  \centering
    \includegraphics[width=0.8\textwidth]{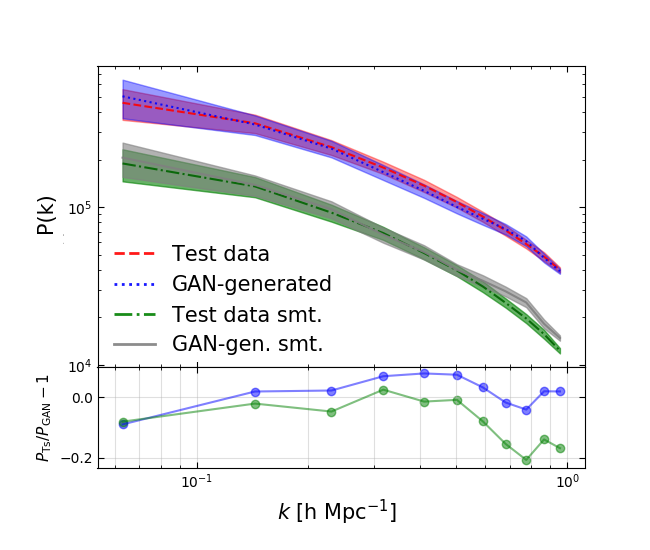}
    \label{cw_fR_plots:3}
  \end{subfigure}
    \begin{subfigure}[b]{0.45\textwidth}
  \centering
    \includegraphics[width=0.8\textwidth]{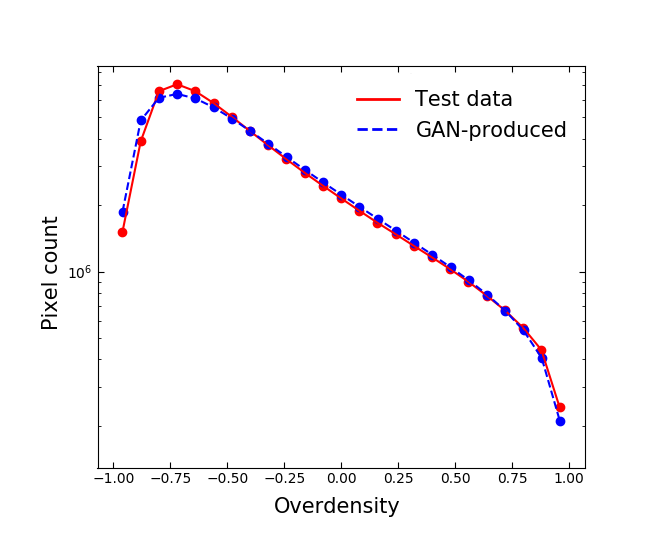}
    \label{cw_fR_plots:4}
  \end{subfigure}
  \caption{A selection of diagnostics to compare the test and the GAN-produced cosmic web slices for $f_{R0} = \{10^{-7}, 10^{-1} \}$ (with $\sigma_{8} = 0.8$ and $z = 0.0$). \textbf{Top left}: power spectra for an ensemble of 64 randomly chosen slices for both datasets; \textbf{top right}: mean and standard deviation of the power spectra produced using 1000 randomly chosen slices with $f_{R0} = 10^{-1}$; \textbf{bottom left}: same as top right, but for $f_{R0} = 10^{-7}$; \textbf{bottom right}: the overdensity histogram (no smoothing). The blue and the green dots give $P_{\rm Ts}/P_{\rm GAN} - 1$ with and without Gaussian smoothing applied correspondingly.  }
  \label{cw_fR_plots}
\end{figure*}

Fig. \ref{figure9} shows the Minkowski functional analysis. In this case, very little deviation is observed. In general, there is a good agreement between the GAN-produced and the training and the test datasets, especially for the first and the second Minkowski functionals. For the third Minkowski functional, the results diverge around the lower trough area, which is also observed for other datasets. This is at least in part related to small-scale noise as indicated by the previous analysis.

\begin{figure*}
\centering
\captionsetup[subfigure]{justification=centering}
  \begin{subfigure}[b]{0.325\textwidth}
  \centering
    \includegraphics[width=0.8\textwidth]{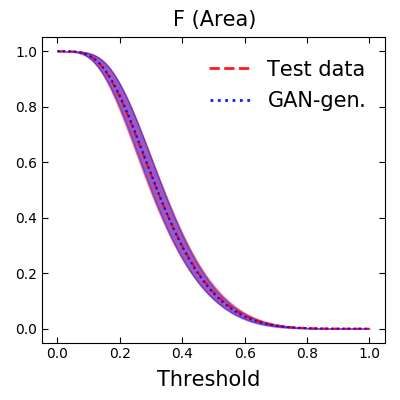}
    \label{cw_fR_MF:1}
  \end{subfigure}
  \begin{subfigure}[b]{0.325\textwidth}
  \centering
    \includegraphics[width=0.8\textwidth]{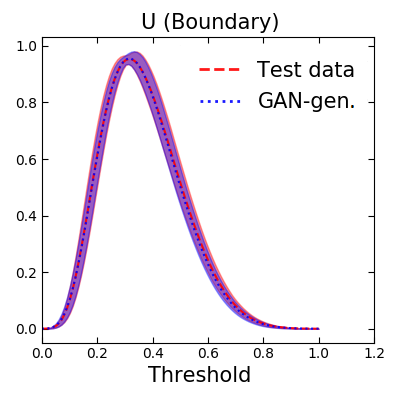}
    \label{cw_fR_MF:2}
  \end{subfigure}
    \begin{subfigure}[b]{0.325\textwidth}
  \centering
    \includegraphics[width=0.8\textwidth]{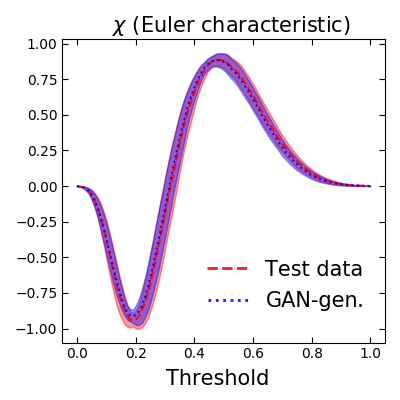}
    \label{cw_fR_MF:3}
  \end{subfigure}
  \caption{A comparison of the Minkowski functionals evaluated using 1000 randomly selected cosmic web slices from the dataset with two different values of $f_{R0} = \{10^{-7},10^{-1}\}$. Gaussian smoothing is applied for both datasets. }
  \label{cw_fR_MF}
\end{figure*}

The results are similar for the GAN trained on cosmic web slices corresponding to different $f(R)$ models (fig. \ref{cw_fR_MF}). In general, we found a good agreement between the datasets, given the standard deviation of the data and the GAN-produced results. Gaussian smoothing, in this case, was more effective in reducing some of the offset observed in the power spectrum analysis. However, it increased the offset on the smallest scales.

\subsection{Dark Matter, Gas and Internal Energy Results}

In the case of training the GAN algorithm on multiple components at the same time, we found the training procedure to be relatively quick and efficient (around 1.3 time quicker compared to the datasets discussed previously) despite the training dataset being 3 times bigger. This is most likely due to the fact that the cosmic web slices in this particular dataset corresponded to a much larger simulation box and hence were not as detailed on the smallest scales. 

\begin{figure*}
\centering
\captionsetup[subfigure]{justification=centering}
  \begin{subfigure}[b]{0.48\textwidth}
  \centering
    \includegraphics[width=0.8\textwidth]{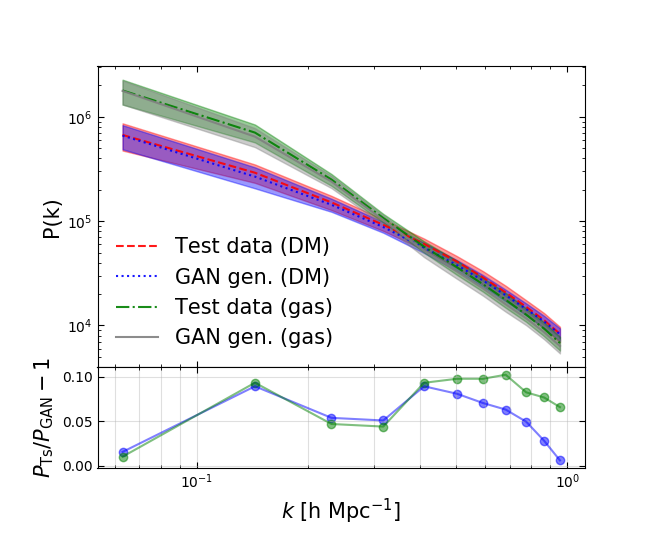}
    \label{cw_illustris:1}
  \end{subfigure}
  \begin{subfigure}[b]{0.48\textwidth}
  \centering
    \includegraphics[width=0.8\textwidth]{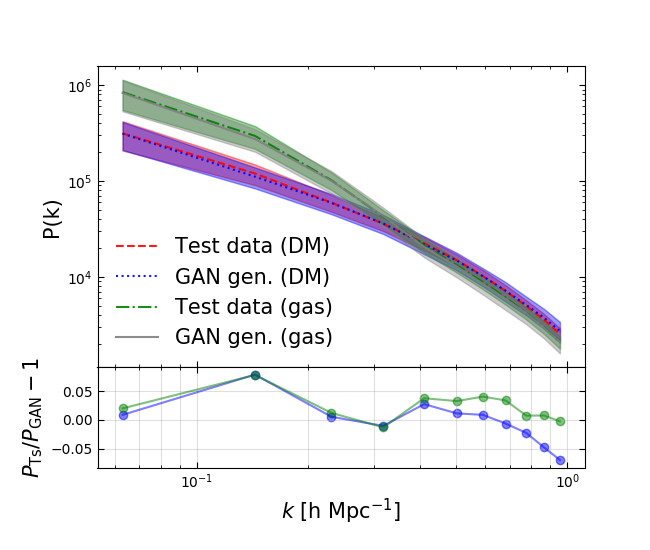}
    \label{cw_illustris:2}
  \end{subfigure}
    \begin{subfigure}[b]{0.48\textwidth}
  \centering
    \includegraphics[width=0.8\textwidth]{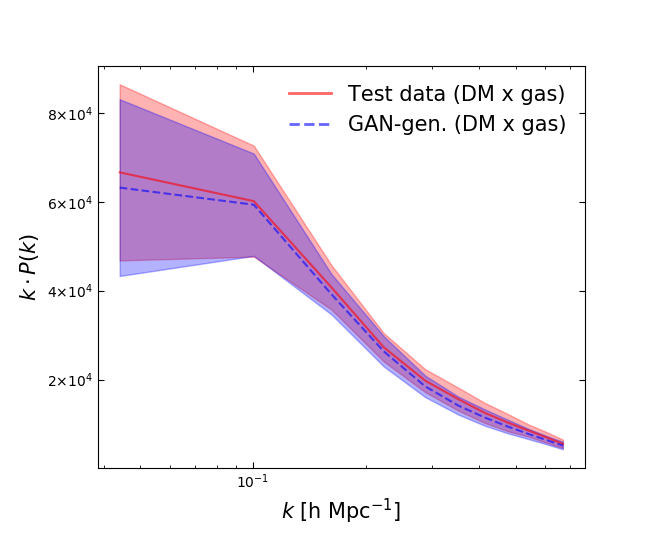}
    \label{cw_illustris:3}
  \end{subfigure}
    \begin{subfigure}[b]{0.48\textwidth}
  \centering
    \includegraphics[width=0.8\textwidth]{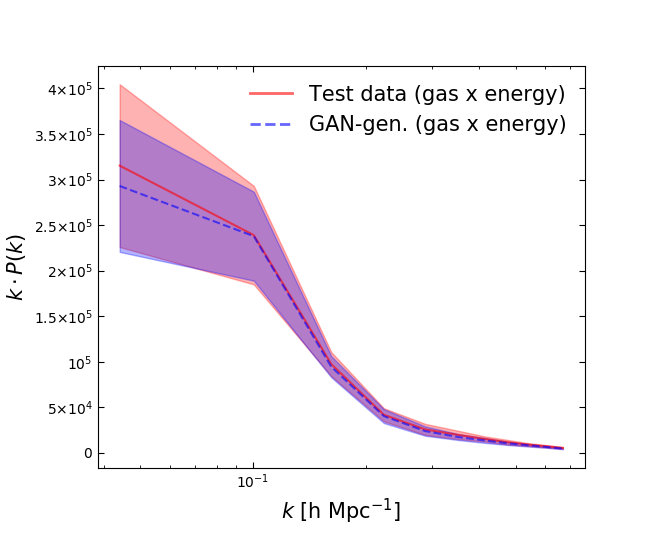}
    \label{cw_illustris:4}
  \end{subfigure}
  \caption{A selection of diagnostics to compare the test and the GAN-produced multi-component cosmic web slices. \textbf{Top left}: the mean and the standard deviation of the power spectrum for 1000 randomly chosen slices for both datasets along with the corresponding relative difference between the datasets (green for $P_{\rm Ts}^{\rm gas} / P_{\rm GAN}^{\rm gas} - 1$ and blue for $P_{\rm Ts}^{\rm DM} / P_{\rm GAN}^{\rm DM} - 1$); \textbf{top right}: same as top left, but with Gaussian smoothing applied; \textbf{bottom left}: the cross-power spectrum calculated between 1000 randomly chosen dark matter and the corresponding gas cosmic web pairs for both the test and the GAN-produced datasets; \textbf{bottom right}: same as bottom left, but for the gas-energy cross-power. }
  \label{cw_illustris}
\end{figure*}

As before, we calculated the relative difference between the GAN-produced and the training and test datasets. The internal energy slices were analysed using Minkowski functionals as well as the cross-power spectrum (fig. \ref{cw_illustris}). The analysis was done for both dark matter and the gas components. The relative difference between the power spectra for both DM and gas cosmic web slices was found to be at around 5\% level for all the covered range. Gaussian smoothing reduced this value to 1-5\%. In addition, the cross-power spectrum was calculated for all the components. For both the dark matter-gas and the gas-energy pairs there is a good agreement between the test and the GAN-produced datasets given the large standard deviation. Both plots show values well above zero for most $k$ values, indicating a significant correlation between the dark matter and the corresponding gas as well as the internal energy distributions on all scales as expected. 

\begin{figure*}
\centering
\captionsetup[subfigure]{justification=centering}
  \begin{subfigure}[b]{0.325\textwidth}
  \centering
    \includegraphics[width=0.8\textwidth]{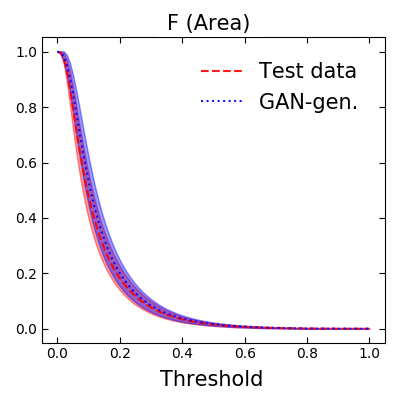}
    \label{cw_dm_gas_MFs:1}
  \end{subfigure}
  \begin{subfigure}[b]{0.325\textwidth}
  \centering
    \includegraphics[width=0.8\textwidth]{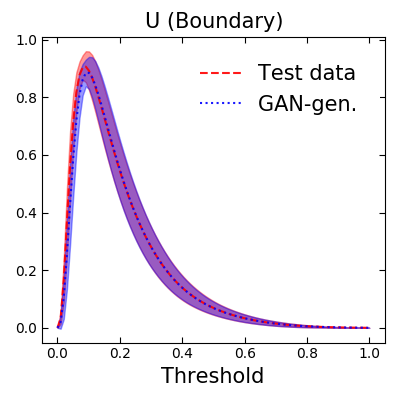}
    \label{cw_dm_gas_MFs:2}
  \end{subfigure}
    \begin{subfigure}[b]{0.325\textwidth}
    \centering
    \includegraphics[width=0.8\textwidth]{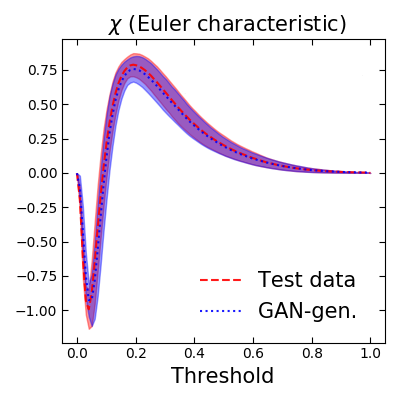}
    \label{cw_dm_gas_MFs:3}
  \end{subfigure}
    \begin{subfigure}[b]{0.325\textwidth}
    \centering
    \includegraphics[width=0.8\textwidth]{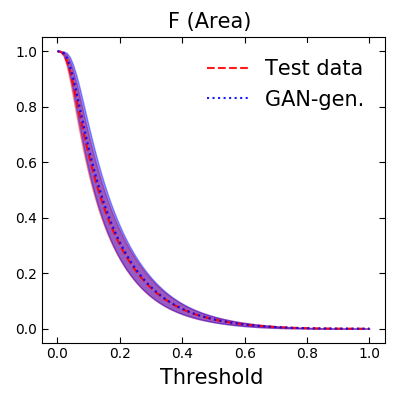}
    \label{cw_dm_gas_MFs:4}
  \end{subfigure}
    \begin{subfigure}[b]{0.325\textwidth}
    \centering
    \includegraphics[width=0.8\textwidth]{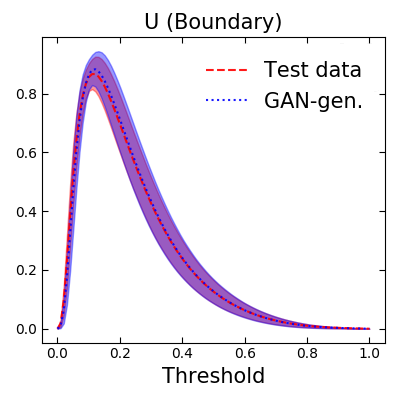}
    \label{cw_dm_gas_MFs:5}
  \end{subfigure}
    \begin{subfigure}[b]{0.325\textwidth}
    \centering
    \includegraphics[width=0.8\textwidth]{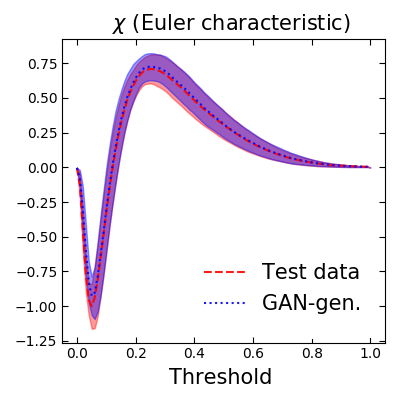}
    \label{cw_dm_gas_MFs:6}
  \end{subfigure}  
    \begin{subfigure}[b]{0.325\textwidth}
    \centering
    \includegraphics[width=0.8\textwidth]{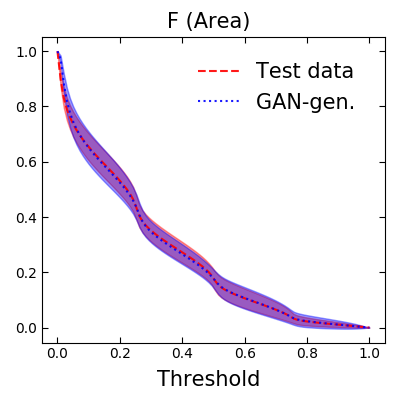}
    \label{cw_dm_gas_MFs:7}
  \end{subfigure}  
    \begin{subfigure}[b]{0.327\textwidth}
    \centering
    \includegraphics[width=0.833\textwidth]{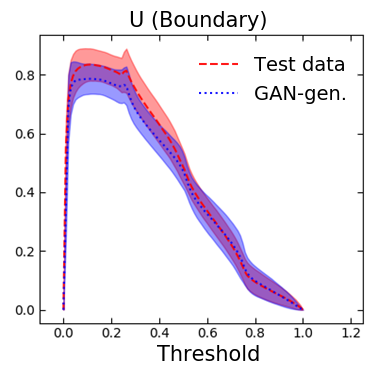}
    \label{cw_dm_gas_MFs:8}
  \end{subfigure}  
    \begin{subfigure}[b]{0.325\textwidth}
    \centering
    \includegraphics[width=0.8\textwidth]{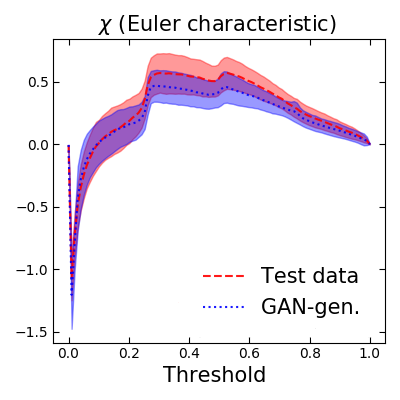}
    \label{cw_dm_gas_MFs:9}
  \end{subfigure} 
  \caption{Results of the Minkowski functional analysis for the GAN trained on the DM, gas and the internal energy data. \textbf{Top row:} Minkowski functionals for the DM cosmic web slices; \textbf{middle row:} Minkowski functionals for the gas overdensity slice data; \textbf{bottom row:} the corresponding Minkowski functionals for the internal energy data. In all cases Gaussian smoothing is applied. }
  \label{cw_dm_gas_MFs}
\end{figure*}

The Minkowski functional analysis (fig. \ref{cw_dm_gas_MFs}) revealed a generally good agreement between the two datasets, with significant differences appearing only in the boundary and the Euler characteristic Minkowski functionals for the energy cosmic web slices. This is somewhat surprising as the internal energy slices in general are significantly less complex on the smallest of scales when compared to the corresponding dark matter and gas data (see fig. \ref{figure3}), hence we expected the GAN to easily learn to reproduce the named dataset. However, we also found that the internal energy data and the corresponding Minkowski functionals are especially sensitive to adding any small scale artificial noise. A more detailed Minkowski functional analysis is required to determine the reason for this divergence.    

\subsection{Latent Space Interpolation Results}
\label{section:latent_results}

To perform the latent space interpolation procedure we trained the GAN to produce cosmic web slices of two different redshifts along with weak lensing maps of different $\sigma_{8}$ values. Once trained, we produced a batch of outputs and in each case chose a pair of slices/maps corresponding to different redshifts or $\sigma_{8}$ values. Subsequently, we interpolated between the input vectors $Z_{1}$ and $Z_{2}$ corresponding to the outputs with different redshifts and $\sigma_{8}$ values (see fig. \ref{figure 2}).   

\begin{figure*}
\centering
\captionsetup[subfigure]{justification=centering}
  \begin{subfigure}[b]{0.42\textwidth}
    \includegraphics[width=0.9\textwidth]{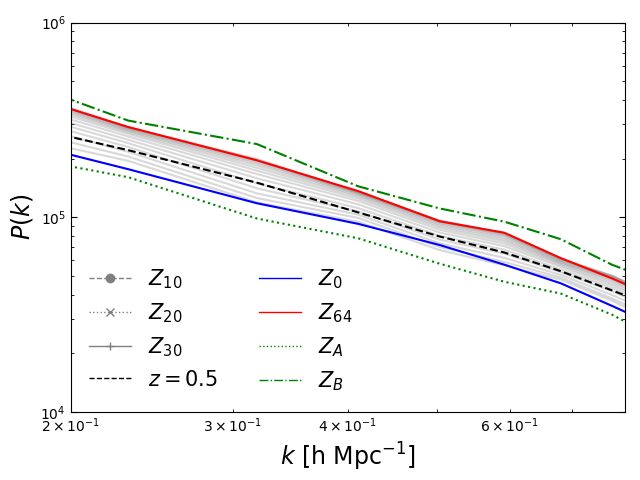}
    \caption{CW slice redshift interpolation}
    \label{fig10:a1}
    
  \end{subfigure}
  \begin{subfigure}[b]{0.42\textwidth}
  \centering
    \includegraphics[width=0.9\textwidth]{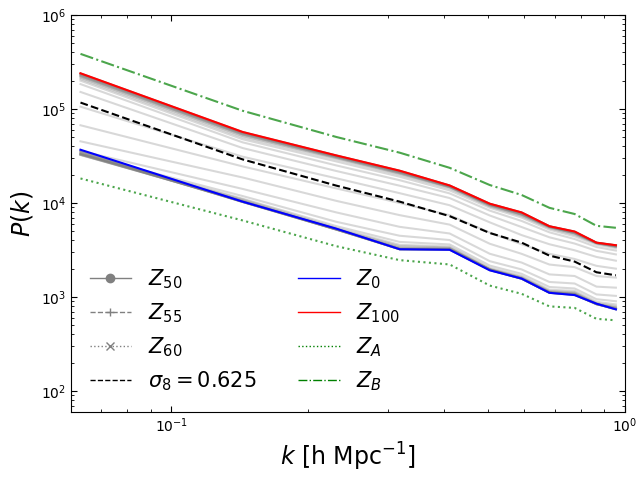}
    \caption{WL $\sigma_{8}$ interpolation}
    \label{fig10:a2}
  \end{subfigure}
  \caption{The results of the linear latent space interpolation technique. \textbf{Left:} the matter power spectrum corresponding to a linear interpolation between two cosmic web slices of redshifts $z=0.0$ and $z=1.0$. The lines in grey are a selection of intermediate output slices generated by the procedure, while the black line corresponds to the mean value of the power spectrum calculated by choosing 100 random (test data) slices of redshift $z=0.5$. \textbf{Right:} interpolating between two randomly chosen weak lensing maps with different values of $\sigma_{8}$. As before, the black line corresponds to the mean power spectrum produced from 100 random maps with $\sigma_{8} = 0.625$.The grey lines were produced by sampling 64 and 100 latent space points correspondingly. Lines denoted by $Z_{A}$ and $Z_{B}$ indicate the power spectra resulting from interpolating beyond the $Z_{0}$ and $Z_{64}/Z_{100}$ points by an amount of $dZ = Z_{2}-Z_{1}$.
  }
  \label{figure10}
\end{figure*}

Fig. \ref{figure10} illustrates the results of the latent space interpolation procedure. In particular, it shows that the technique does indeed produce intermediate power spectra. However, the transition is not linear -- the power spectra lines corresponding to equally spaced inputs (in the latent space) are not equally spaced in the power spectrum space. This is the case as the produced data samples can be described as points on a Riemannian manifold, which in general has curvature (see appendix \ref{appendix:riemannian_geometry} for more details).

\begin{figure*}
  \centering
    \includegraphics[width=0.88\textwidth]{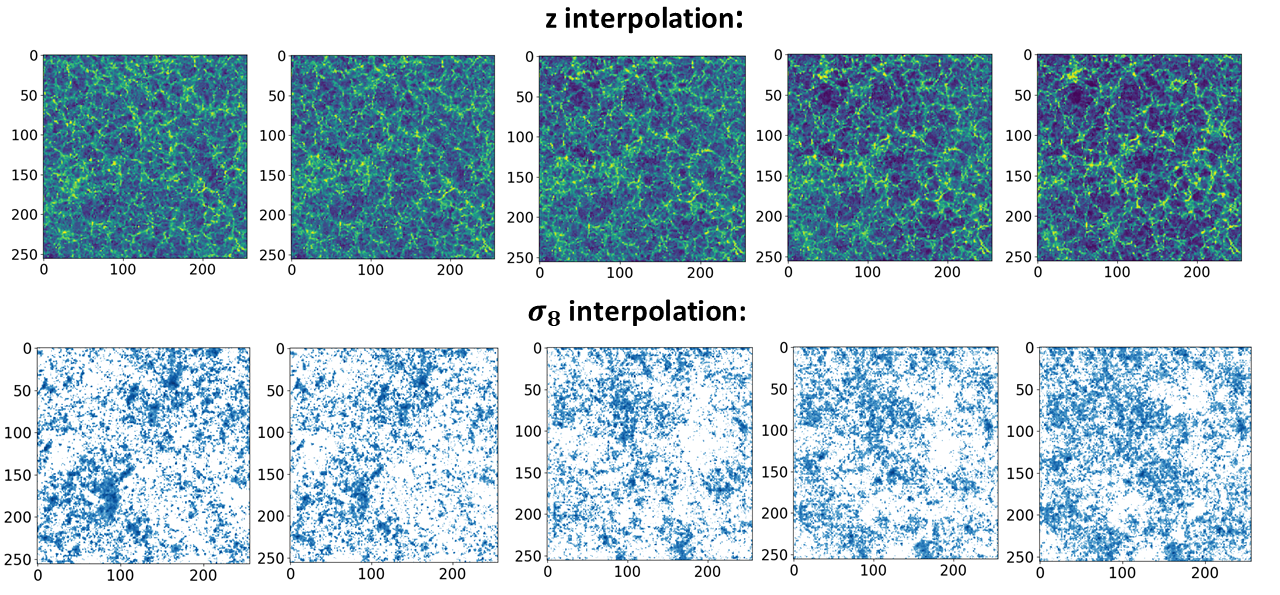}
    \caption{The results of the latent space interpolation procedure for cosmic web slices of redshifts $z=0.0$ (far right) and $z=1.0$ (far left) and weak lensing convergence maps of $\sigma_{8}^{1} = 0.436$ (far left) and $\sigma_{8}^{1} = 0.814$ (far right).  }
    \label{figure11}
\end{figure*}

Fig. \ref{figure10} and \ref{figure11} show the results of interpolating between cosmic web slices with redshifts $z=0.0$ and $z=1.0$ and weak lensing maps with $\sigma_{8} =0.436$ and $\sigma_{8} = 0.814$. The interpolated samples are statistically realistic and the transition is nearly smooth.

An important part of the latent space interpolation procedure is being able to distinguish between the GAN-generated cosmic web slices and weak lensing maps of different redshifts, cosmologies and modified gravity parameters. In this regard, we have used tested two machine learning algorithms: a convolutional neural network and gradient boosted decision trees. We initially used a 3-layer convolutional neural network with 128 and 64 output filters of the convolution correspondingly with kernel size equal to  $3 \times 3$ px and \textit{tanh} activation functions. We found that the neural network approach has mostly failed to distinguish between the different dataset classes reliably. After a thorough hyperparameter search we managed to reach accuracy of around 75\%, which was not good enough for the given task. We found that the gradient boosted decision tree algorithm (\textit{XGBoost} \citep{chen2016_xgboost}) was fast and accurate when predicting the dataset class. In particular, we reached 95-98\% accuracy (depending on the dataset and hyperparameters used), when predicting the dataset class of unseen test samples. Table \ref{table:XGBoost_parameters} summarizes the parameters used when training the \textit{XGBoost} algorithm. 

\begin{table}
\centering
\begin{tabular}{ |c||c|c|c|c| } 
 \hline
 \textbf{Parameter:} & $R^{\rm XGB}_{\rm L}$ & $D_{max}$ & Training step & Objective\\ 
 \hline 
 \textbf{Value:} & 0.08 & 2  & 0.3  & \textit{multi:softprob}\\ 
 \hline
\end{tabular}
\caption{The \textit{XGBoost} parameters used for classifying the cosmic web slices with redshifts $z=\{0.0,1.0\}$ and the weak lensing maps with $\sigma_{8} = \{ 0.436,0.814 \}$. $R^{\rm XGB}_{\rm L}$ refers to the learning rate and $D_{max}$ is the maximum tree depth.   }
\label{table:XGBoost_parameters}
\end{table}

Combining such a machine learning approach with a power spectrum analysis allowed us to distinguish between the different classes of the GAN-produced outputs reliably.

The latent space interpolation results illustrate a number of interesting features of GANs. Firstly, the results illustrate that the GAN training procedure tightly encodes the various features discovered in our training dataset in the high-dimensional latent space. By finding clusters in this latent space, corresponding to outputs of different redshifts or cosmology parameters, and linearly interpolating between them, we can produce outputs with intermediate values of the mentioned parameters. This allows us to indirectly control the outputs produced by the generator. In addition, as illustrated by fig. \ref{figure10}, it is possible to discover vectors in the latent space that correspond to changes in redshift and $\sigma_{8}$. More generally, different vectors in the latent space correspond to various features of the dataset, for instance the shapes of the filaments and the voids in the case of the cosmic web slices. This illustrates that the latent space interpolation technique is crucial for controling the outputs of the GAN algorithm as well as accessing information about the important dataset features learnt by the GAN during the training procedure. It is important to note, however, that these features are known to be entangled, such that sampling input points along some vector in the latent space inevitably introduces multiple changes to the generated outputs (see section \ref{conclusions} for a further discussion of this issue).

\section{Analysis and Conclusions}
\label{conclusions}
The main goal of this work was to investigate whether GANs can be used as a fast and efficient emulator capable of producing realistic and novel mock data. Likewise, we investigated the model feature space using the technique of latent space interpolation. Our results are encouraging, illustrating that GANs are indeed capable of producing realistic mock datasets. In addition, we have shown that GANs can be used to emulate dark matter, gas and internal energy distribution data simultaneously. This is a key result, as generating realistic gas distributions requires complex and computationally expensive hydrodynamical simulations. Hence, producing vast amounts of realistic multi-component mock data quickly and efficiently will be of special importance in the context of upcoming observational surveys.    

The GAN-produced data in general cannot be distinguished from the training dataset visually. In terms of the power spectrum analysis, the relative difference between the GAN-produced and the test data ranges between 1-20\%  depending on the dataset and whether Gaussian smoothing was applied. The Minkowski functional analysis revealed a generally good agreement between the two datasets with an exception of the third Minkowski functional corresponding to curvature, which showed subtle differences for all studied datasets. In addition, greater differences were observed when training the GAN on datasets with multiple data classes. This is somewhat expected, as the training task becomes more difficult. In general, when training on datasets with multiple redshifts and cosmological parameters, our algorithm is capable of separating different datasets. It is important to emphasize, however, that this depends significantly on the dataset used and on how different the cosmological parameters are. For datasets with significant overlap in terms of parameters of interest, an alternative approach might be required. In addition, it is important to note that the observed differences between the GAN-generated and the test data can be partially accounted for as a result of the small-scale noise in the emulated images. We found Gaussian smoothing with a $3 \times 3$ pixel kernel size to be effective in filtering away most of such noise. Finally, we note that the training datasets used in this work are smaller than those used in \cite{rodriguez2018, Mustafa2019}, which, at least partially, accounts for the differences between our and their corresponding results.

We also investigated a commonly used technique of latent space interpolation as a tool for investigating the feature space of the model. Interestingly, we found that such a procedure allows us to generate samples with intermediate redshift/cosmology/$f_{R0}$ parameter values, even if our model had not been explicitly trained on those particular values. In addition, the latent space interpolation procedure offers an indirect way of controlling the outputs of the GAN. However, it is important to point out some of the drawbacks of this procedure. Namely, as pointed out in machine learning literature, the latent space of a convolutional GAN is known to be entangled. In other words, moving in a different direction in the latent space necessarily causes multiple changes to the outputs of the GAN. As a concrete example, finding a latent space line that induces a change in redshift of a given output necessarily also introduces other subtle changes to the output (e.g. the depth of the voids or the distribution of the filaments). So if we take a random output of redshift $z = 1.0$ and perform the linear interpolation procedure to obtain a cosmic web slice of $z = 0.0$, the obtained slice will correspond to a realistic but \textit{different} distribution of the required redshift. This is a drawback as in an ideal case we would love to have full control of individual parameters, while not affecting other independent features of a dataset. There are however other generative models discussed in the literature that allow such manipulation of the latent space. Namely, the $\beta$-VAE variational autoencoder and the InfoGAN algorithms, allow encoding features into the latent space in a special way that allows full control of individual key parameters without affecting the other features of the dataset (latent space disentanglement) \citep{chen2016, higgins2017, burgess2018}. 

As illustrated by our results, the technique of latent space interpolation can be used to indirectly control the outputs of the GAN algorithm. A significant disadvantage of such approach is that it is based on the ability to find certain vectors in the latent space, which, given its high dimensionality, is not trivial. In addition, it is difficult to tell apart outputs corresponding to different cosmological parameters (hence the need for an extra machine learning classifier). These issues, however, can be tackled by directly specifying the cosmological parameters as an extra input vector alongside the usual latent space input. Such an approach is applied in the recent GAN architectures such as the mentioned conditional GAN algorithm \citep{Mirza2014,Perraudin2020}. The conditional GAN architecture also makes it easier to interpolate between the outputs corresponding to different cosmological parameters. Nonetheless, even though the code used in this work is not the ideal tool for directly controlling the GAN outputs, it holds multiple advantages when training on a single-parameter datasets. Namely, our open-source code (see the data availability section below) is well tested and easily customizable. In addition, as illustrated by our Illustris dataset results, our code allows training multiple data components (e.g. dark matter, gas and energy distribution data) simultaneously with no difficulty. This makes it particularly easy to apply the techniques laid out in this work on other kinds of datasets. Lastly, it should be noted that the latent space interpolation techniques presented here are architecture-independent and thus can be easily integrated into newer GAN architectures.

Another important topic to discuss is the problem of mode collapse. As is widely discussed in the literature, the generator neural network is prone to getting stuck in producing a very small subsample of realistic mock datapoints that fool the discriminator neural network. Resolving mode collapse is an important open problem in the field of deep learning, with a variety of known strategies ranging from choosing a particular GAN architecture, to altering the training procedure or the cost function \citep{srivastava2017,hong2019}. Mode collapse was encountered multiple times in our training procedure as well. As a rule of thumb, we found that reducing the learning rate parameter had the biggest effect towards resolving mode collapse for all studied datasets. Learning rates around the values of $3 \times 10^ {-5}$ for the cosmic web data and $9 \times 10^{-6}$ for the weak lensing maps were found to be the most effective in avoiding any mode collapse. 

As we have shown, GANs can be used to generate novel 2-D data efficiently. A natural question to ask is whether this also applies to 3-D data. As an example, an analogous emulator capable of generating 3-D cosmic web data, such as that produced by state of the art hydrodynamic and DM-only simulations would be very useful. In principle there is no limit on the dimensionality of the data used for training a GAN, however, in practice, going from 2-D to 3-D data leads to a significant increase of the generator and the discriminator networks. In addition, in the case of 3-D cosmic web data, forming a big enough training dataset would become an issue, as running thousands of simulations would be required. However, as previously mentioned, there are sophisticated ways of emulating 3-D cosmic web data as shown in \cite{perraudin2019}, where a system of GANs is used to upscale small resolution comic web cubes to full size simulation boxes. Note that the techniques introduced in this work (e.g. latent space interpolation) can be readily combined with the mentioned 3-D approach. 

A number of other interesting directions can be explored in future work. Namely, a more detailed investigation into the Riemannian geometry of GANs could lead to a better understanding of the feature space of the algorithm. Finally, many other datasets could be explored. With upcoming surveys, such as Euclid, generating mock galaxy and galaxy cluster data quickly and efficiently is of special interest. A GAN could be used to generate galaxies with realistic intrinsic alignments, density distributions and other properties. Similarly, GANs could be used to quickly emulate realistic galaxy cluster density distributions at a fraction of the computational cost required to run full hydrodynamic simulations. 

To conclude, GANs offer an entirely new approach for cosmological data emulation. Such a game theory based approach has been demonstrated to offer a quick and efficient way of producing novel data for a low computational cost. As we have shown in this work, the trade-off for this is a 1-20\% difference in the power spectrum, which can be satisfactory or not depending on what application such an emulator is used for. Even though a number of questions remain to be answered regarding the stability of the training procedure and training on higher dimensional data, GANs will undoubtedly be a useful tool for emulating cosmological data in the era of modern \textit{N}-body simulations and precision cosmology.  

\subsection*{Data Availability}
\label{data_code_availability}

The key datasets generated in this work as well as the main analysis scripts are available at following GitHub repository: \href{https://github.com/AndriusT/cw_wl_GAN}{https://github.com/AndriusT/cw\_wl\_GAN}. The link also contains detailed instructions on how to produce the data samples from the publicly available Illustris data. The full Illustris datasets can be found at: \href{https://www.illustris-project.org/data/}{https://www.illustris-project.org/data/}.

The used weak lensing datasets with different cosmological parameters can be accessed at: \href{http://columbialensing.org/}{http://columbialensing.org/}.

\subsection*{Acknowledgments}

We thank the Columbia Lensing group (\href{http://columbialensing.org}{http://columbialensing.org}) for making their suite of simulated maps available, and NSF for supporting the creation of those maps through grant AST-1210877 and XSEDE allocation AST-140041. 

This work would also not have been possible without the access to the HPC Sciama facilities at the Institute of Cosmology and Gravitation and the support by the IT staff. 

We also thank Adam Amara for the guidance and the useful discussions. Finally, we thank Minas Karamanis for the help with the statistical analysis.    

AT is supported by the Science \& Technology Facilities Council (STFC) through the DISCnet Centre for Doctoral Training funding. The Google Cloud computational resources were funded by the Research and Innovation Funding by the University of Portsmouth. 

KK has received funding from the European Research Council (ERC) under the European Union's Horizon 2020 research and innovation programme (grant agreement No. 646702 "CosTesGrav"). DB and KK are also supported by the UK STFC ST/S000550/1.



\bibliographystyle{mnras}
\bibliography{refs} 




\appendix

\section{Samples of the GAN-produced Data}
\label{samples}

\begin{figure*}
  \centering
    \includegraphics[width=0.80\textwidth]{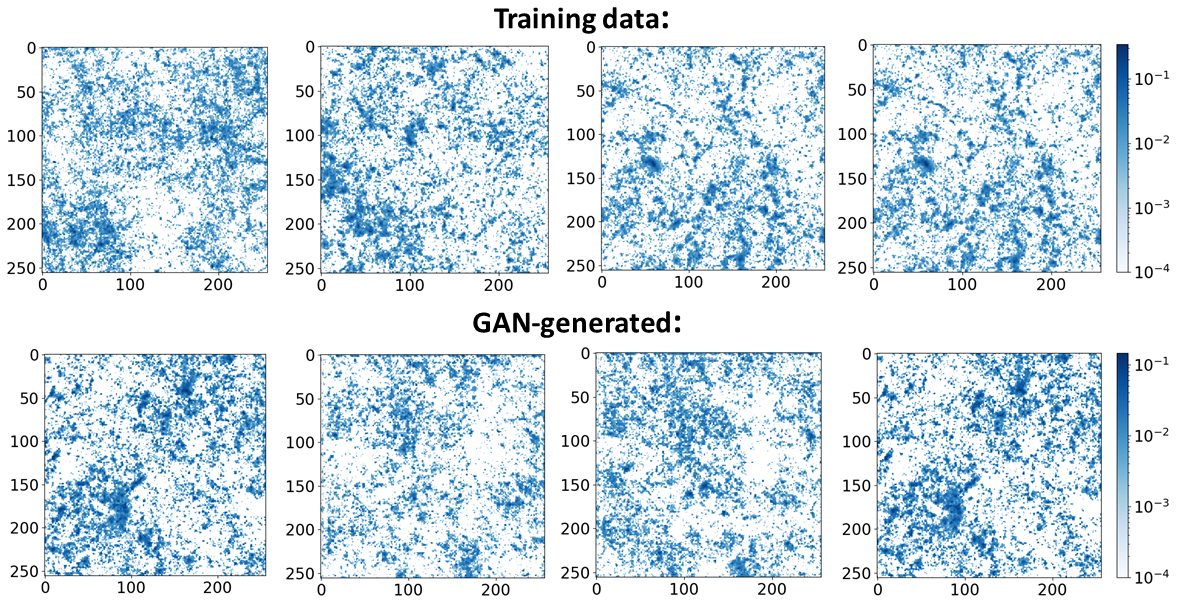}
    \caption{ A comparison of 4 randomly selected weak lensing convergence maps. The colors are log-normalized to emphasize the main features and to allow a direct comparison with the previous results in the literature.}
    \label{cosmoGAN_samples}
\end{figure*}

This section contains a selection of GAN-produced samples for visual inspection. Fig. \ref{cosmoGAN_samples} contains randomly selected weak lensing convergence maps produced by the GAN algorithm (these are the samples described in sections \ref{section_weak_lensing_multiple_cosmologies} and \ref{section_cosmic_web_redshifts}). 

\begin{figure*}
  \centering
    \includegraphics[width=0.8\textwidth]{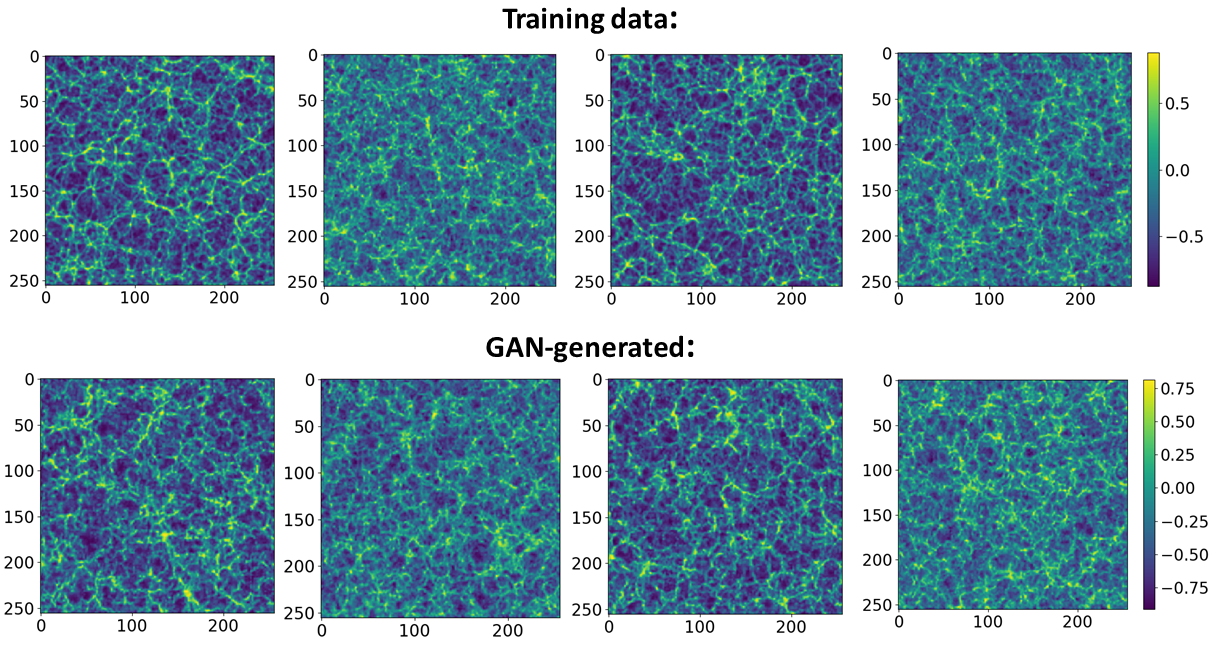}
    \caption{ A comparison of 4 randomly selected cosmic web slices. Columns 1 and 3 correspond to redshift 0.0 while columns 2 and 4 are redshift 1.0.}
    \label{cosmoGAN_cw_samples}
\end{figure*}

Fig. \ref{cosmoGAN_cw_samples} shows a selection of randomly selected cosmic web 2-D slices for two different redshifts. Both the training data and the produced slices have been Gaussian-smoothed. 

\section{Riemannian Geometry of the GAN Algorithm}
\label{appendix:riemannian_geometry}

Recently various connections between GANs and Riemannian geometry have been explored in the machine learning literature. Such connections are important to explore, not only for the sake of curiosity, but also because they allow us to describe GANs and their optimization procedure in a language more familiar to physicists. A Riemannian geometry description of GANs is also powerful when exploring the latent space of a trained generator neural network and the outputs that it produces. Finally, a differential geometry description could shine some light on the connections between generative models and information geometry, which is a well-established field and could offer some new insights into training and analyzing the outputs of such models.

\begin{figure*}
  \centering
    \includegraphics[width=0.50\textwidth]{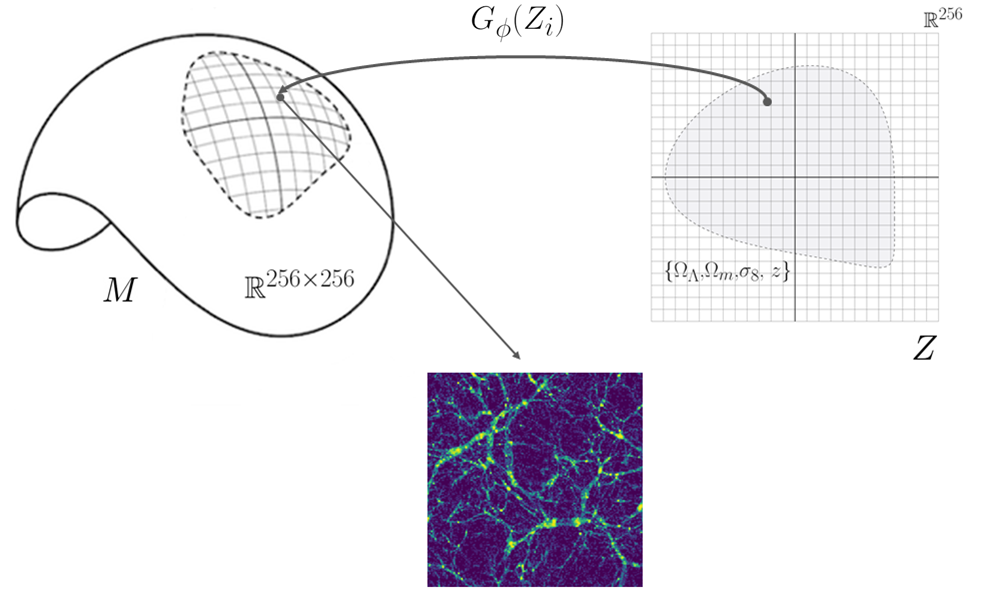}
    \caption{Riemannian geometry of generative adversarial networks. The generator $G_{\phi}(Z_{i})$ can be treated as a mapping from the lower dimensional Euclidean latent space $Z$ (corresponding to the random noise input) to the high dimensional data (pixel) space $M$ (in general non-Euclidean). Each point on $M$ corresponds to a weak lensing map (or a cosmic web slice). }
    \label{riemanian_geometry}
\end{figure*}

Recent work in \cite{shao2017} proposes treating the trained generator neural network as a mapping from a lower dimensional latent space $Z$ to the higher dimensional data space $X$: $G_{\phi}: Z \rightarrow X$ (see fig. \ref{riemanian_geometry}). More specifically, the generator $G_{\phi}(Z)$ maps the latent space vectors of size $n$ (in our case $n = 256$ or $64$) to a manifold $M$ of dimensionality $m$ ($256 \times 256$, i.e. the number of pixels in the output images). Manifold $M$ here simply refers to a subset of the data space (all possible combinations of pixel values), which correspond to realistic images of weak lensing / cosmic web slices. The existence of such a manifold is postulated by the \textit{manifold hypothesis} in deep learning, which states that high-dimensional data can be encoded on a manifold of a much lower dimension. 

Hence if we treat the generator neural network $G_{\phi}$ as a mapping for the latent space to the data space manifold, one can naturally define an \textit{induced metric} $g$, which is simply a product of the Jacobian and the transposed Jacobian: 

\begin{equation}
    g = J(Z)^{T}J(Z) 
\end{equation}

\noindent The Jacobian in our case simply refers to the partial derivative of each output value w.r.t. to each input value, i.e.: 

\[
J = \begin{bmatrix} 
    \frac{\partial X^{1}}{\partial Z^{1}} & \frac{\partial X^{1}}{\partial Z^{2}} & \dots  &\frac{\partial X^{1}}{\partial Z^{n}} \\
    \vdots & \vdots &\ddots & \\
    \frac{\partial X^{m}}{\partial Z^{1}}& \frac{\partial X^{m}}{\partial Z^{2}} & \dots        & \frac{\partial X^{m}}{\partial Z^{n}} 
    \end{bmatrix}
\]

\noindent Once a metric is defined, we can use the usual tools to describe geodesics on the manifold $M$. For instance, we can define a curve $\kappa$ between two points $a$ and $b$ in the latent space $Z$ parametrized by some parameter $t$. Using the mapping $G_{\phi}$, the corresponding curve on the manifold $M$ is then: $G_{\phi}(\kappa(t)) \in M$. To find a curve that corresponds to a geodesic on the manifold, one has to solve the Euler-Lagrange equation: 

\begin{equation}
    \frac{d^{2}\kappa^{\alpha}}{dt^{2}} = - \Gamma^{\alpha}_{\beta \gamma} \frac{d\kappa^{\beta}}{dt} \frac{d\kappa^{\gamma}}{dt},
\end{equation}

\noindent where $\Gamma$ is the usual Christoffel symbol, given by:

\begin{equation}
    \Gamma^{\alpha}_{\beta \gamma} = \frac{1}{2} g^{\alpha \delta} \bigg(\frac{\partial g_{\delta \beta}}{\partial X^{\gamma}} + \frac{\partial g_{\delta \gamma}}{\partial X^{\alpha}} - \frac{\partial g_{\alpha \beta}}{\partial X^{\delta}} \bigg).
\end{equation}

As discussed in \cite{shao2017} geodesics between points on the manifold are of special importance, as they give the smoothest possible transition between multiple outputs. One of the main findings in \cite{shao2017} was that the Riemannian curvature of the manifold corresponding to the their data was surprisingly small and, hence, linear interpolation produced realistic results comparable to the results produced by calculating a geodesic curve between outputs. In our work we also found that linear interpolation generally produced realistic results. However, to ensure that the outputs produced via the latent space interpolation are indeed realistic, one would have to interpolate on a curve in the latent space (corresponding to the geodesic connecting the needed outputs on the data manifold $M$) rather than a line.

Another important connection to Riemannian geometry comes in the context of the discriminator neural network. The discriminator can be viewed as a mapping from the data manifold to a probability manifold $P$, where each point on the manifold corresponds to the probability of a given data sample being real (i.e. belonging to the training dataset). Such a manifold looks remarkably similar to the statistical manifolds studied in the field of information geometry. Insights from information geometry have a long tradition of being used in neural network optimization (e.g. \cite{Hauser2017}). Exploring such connections could lead to deeper insights into the GAN training process, which we plan to explore in future work.

\section{Generator and the Discriminator Architecture}
\label{appendix:GAN_architecture}

The architecture used when training the GAN algorithm is summarized in tables \ref{generator_architecture} and \ref{discriminator_architecture}. 

\begin{table}
\centering
\scalebox{0.8}{%
\begin{tabular}{lccc}
\hline & \textbf{Activ.} & \textbf{Output shape} & \textbf{Params.} \\
\hline Latent & $-$ & 64 & $-$ \\
\hline Dense & $-$ & $512 \times 16 \times 16$ & $8.5 \mathrm{M}$ \\
BatchNorm & $\mathrm{ReLU}$ & $512 \times 16 \times 16$ & 1024 \\
\hline TConv $5 \times 5$ & $-$ & $256 \times 32 \times 32$ & $3.3 \mathrm{M}$ \\
BatchNorm & $\mathrm{ReLU}$ & $256 \times 32 \times 32$ & 512 \\
\hline TConv $5 \times 5$ & $-$ & $128 \times 64 \times 64$ & $819 \mathrm{K}$ \\
BatchNorm & $\mathrm{ReLU}$ & $128 \times 64 \times 64$ & 256 \\
\hline TConv $5 \times 5$ & $-$ & $64 \times 128 \times 128$ & $205 \mathrm{K}$ \\
BatchNorm & $\mathrm{ReLU}$ & $64 \times 128 \times 128$ & 128 \\
\hline TConv $5 \times 5$ & Tanh & $1 \times 256 \times 256$ & 1601 \\
\hline \multicolumn{2}{l} { Total trainable parameters } &   &$\mathbf{1 2 . 3 M}$ \\
\hline
\end{tabular}}
\caption{The architecture of the generator neural network. \textit{TConv} corresponds to the transposed convolutional layer with $\mathrm{stride}=2$ (and the kernel size given by the shown numerical values). \textit{ReLU} corresponds to the rectified linear unit activation function. \textit{LReLU} stands for the leaky rectified linear unit activation function with the leakiness parameter $=0.2$.}
\label{generator_architecture}
\end{table}

\begin{table}
\centering
\scalebox{0.8}{%
\begin{tabular}{lccc}
\hline & \textbf{Activ.} & \textbf{Output shape} & \textbf{Params.} \\
\hline Input map & $-$ & $1 \times 256 \times 256$ & $-$ \\
\hline Conv $5 \times 5$ & LReLU & $64 \times 128 \times 128$ & $1664$ \\
\hline Conv $5 \times 5$ & $-$ & $128 \times 64 \times 64$ & $205 \mathrm{K}$ \\
BatchNorm & LReLU & $128 \times 64 \times 64$ & 256 \\
\hline Conv $5 \times 5$ & $-$ & $256 \times 32 \times 32$ & $819 \mathrm{K}$ \\
BatchNorm & LReLU & $256 \times 32 \times 32$ & 512 \\
\hline Conv $5 \times 5$ & $-$ & $512 \times 16 \times 16$ & $3.3 \mathrm{M}$ \\
BatchNorm & LReLU & $512 \times 16 \times 16$ & 1024 \\
\hline Linear& Sigmoid & $1$ & $131 \mathrm{K}$ \\
\hline \multicolumn{2}{l} { Total trainable parameters } &   &$\mathbf{4.4 M}$ \\
\hline
\end{tabular}}
\caption{The architecture of the discriminator neural network. }
\label{discriminator_architecture}
\end{table}

\section{Latent Space Clustering}
\label{appendix:latent_space_clustering}

\begin{figure}
  \centering
    \includegraphics[width=0.35\textwidth]{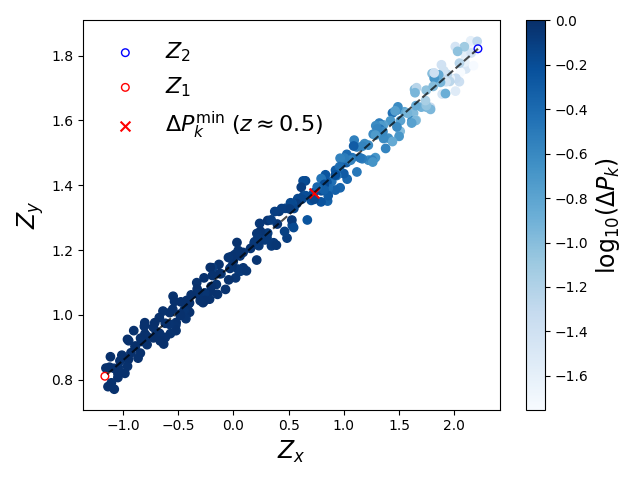}
    \caption{The visualization of the latent space clustering. The points are coloured based on eq. \ref{eq:latent_space_clustering} with the mean power spectrum of 1000 randomly chosen cosmic web slices with redshift $z=0.0$ as the base power spectrum. The axes correspond to two randomly chosen axes out of the 256 dimensions in the latent space. The red x marks the point that produces the power spectrum that is the closest to the mean power spectrum of 1000 cosmic web slices with $z=0.5$ (with the mean difference of 1.4 \%).  }
    \label{latent_space_clustering1}
\end{figure}

To investigate the structure of the latent space two random latent space points were selected corresponding to two cosmic web slices with redshifts $z = \{0.0,1.0 \}$ and weak lensing maps with $\sigma_{8} = \{0.436, 0.814\}$. Then the line connecting the two points was found and a number of points (close to the line, but otherwise randomly distributed) were sampled. For each point, a corresponding output was produced and the power spectrum was calculated. For each point the resulting power spectrum was compared against a base power spectrum (corresponding to $z = 0.0$ and $\sigma_{8} = 0.436$). The comparison was done by calculating the following quantity: 

\begin{equation}
\Delta P_{k} = \sum_{i} |\rm log_{10}(P(k_{i})) - log_{10}(P^{\rm B}(k_{i})| ,
\label{eq:latent_space_clustering}
\end{equation}

\noindent with $P^{B}$ as the base power spectrum. The mentioned statistic was then rescaled in the range of $[0.0,1.0]$ and used to colour the points in figure \ref{latent_space_clustering1}. Note that the dimensionality of the latent space is too high to be fully visualized, hence a 2-D projection was plotted instead (i.e. only 2 of the 256 dimensions were used).

\begin{figure}
  \centering
    \includegraphics[width=0.35\textwidth]{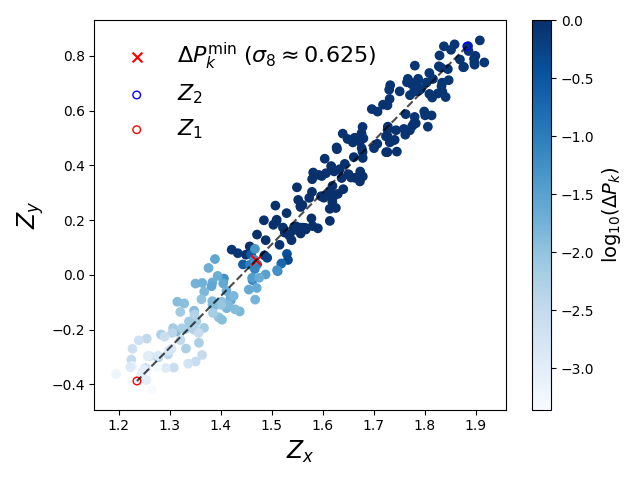}
    \caption{Same as fig. \ref{latent_space_clustering1} except for the weak lensing maps with $\sigma_{8}$ = $\{0.436,0.814\}$. The mean power spectrum of 1000 maps with $\sigma_{8} = 0.436$ is used as the base power spectrum for eq. \ref{eq:latent_space_clustering}. The red x marks the point with the power spectrum closest to the mean power spectrum due to 1000 maps with $\sigma_{8} = 0.625$ (with $2.7\%$ difference). }
    \label{latent_space_clustering2}
\end{figure}

Figures \ref{latent_space_clustering1} and \ref{latent_space_clustering2} illustrates some of the key properties of the latent space of the GAN algorithm. In particular, as expected, points that are nearby in latent space, produce outputs with similar power spectrum. Note that most of the plotted points produce outputs that have power spectra corresponding to either $z=0.0$ or $z=1.0$ or correspondingly to the two $\sigma_{8}$ values for the weak lensing maps. In addition, one can see that the transition between these two regimes is not entirely continuous (this depends partially on the chosen activation functions). Nonetheless, in both cases there exists a cluster of points, near the limit between the two regimes, which produces power spectra nearly identical to $z=0.5$ and $\sigma_{8} = 0.436$. One can always generate more points by zooming in that region in the latent space and sampling extra points.

\label{lastpage}

\end{document}